%% file: preprint.tex
\documentclass[11pt,a4paper]{article}
\usepackage{amsmath}
\usepackage{amssymb}
\usepackage{amsthm}
\input{preamble/preamble}

\makeatletter
\newcommand{\LoadIfMissing}[2][]{\@ifpackageloaded{#2}{}{\usepackage[#1]{#2}}}
\LoadIfMissing[margin=2.5cm]{geometry}
\LoadIfMissing{amsmath}
\LoadIfMissing{amssymb}
\LoadIfMissing{amsthm}
\LoadIfMissing{bm}
\LoadIfMissing{graphicx}
\LoadIfMissing{booktabs}
\LoadIfMissing{algorithm}
\LoadIfMissing{algpseudocode}
\LoadIfMissing{authblk}
\theoremstyle{plain}
\@ifundefined{theorem}{\newtheorem{theorem}{Theorem}}{}
\@ifundefined{lemma}{\newtheorem{lemma}{Lemma}}{}
\@ifundefined{proposition}{\newtheorem{proposition}{Proposition}}{}
\@ifundefined{corollary}{\newtheorem{corollary}{Corollary}}{}
\theoremstyle{definition}
\@ifundefined{definition}{\newtheorem{definition}{Definition}}{}
\@ifundefined{assumption}{\newtheorem{assumption}{Assumption}}{}
\theoremstyle{remark}
\@ifundefined{remark}{\newtheorem{remark}{Remark}}{}
\providecommand{\texorpdfstring}[2]{#1}
\makeatother

\input{acronyms/acronyms}

\title{$\bm{\lambda}$\emph{-learning}: Temporal Difference Learning of the Value Function Gradient for State Estimation and Smoothing}

\author[1]{Stefano Menchetti\thanks{Corresponding author: Stefano Menchetti, IMT School for Advanced Studies Lucca, Piazza San Francesco 19, 55100 Lucca, Italy.}}
\author[2]{Daniele Masti}
\author[1]{Mario Zanon}
\author[1]{Alberto Bemporad}
\affil[1]{IMT School for Advanced Studies Lucca, Piazza San Francesco 19, 55100 Lucca, Italy}
\affil[2]{Gran Sasso Science Institute, Viale Francesco Crispi 7, 67100 L'Aquila, Italy}
\affil[ ]{\texttt{\{stefano.menchetti, mario.zanon, alberto.bemporad\}@imtlucca.it}, \texttt{daniele.masti@gssi.it}}
\date{\small Preprint submitted to \emph{Automatica}.}

\begin{document}

\maketitle

\begin{abstract}
    We develop a gradient-based approach to recursive state estimation and smoothing grounded in the dynamic programming structure of the full information estimation problem. Instead of approximating the value function, we work with its gradient, the $\lambda$-function, and show that it satisfies a forward recursion in which past information enters implicitly through optimality conditions. This replaces bootstrapped value-function targets with a temporal-difference residual on the $\lambda$-function, leading to a learning problem that enforces estimation and smoothing optimality conditions directly from measurement data, without requiring ground-truth states. The learned $\lambda$-function defines both a recursive estimator and a smoother within a unified formulation. For an idealized version of the learning problem, we establish local convergence of gradient descent to a critical point under smoothness and regularity assumptions. A practical semi-gradient algorithm, termed \emph{$\lambda$-learning}, is obtained by combining forward estimation rollouts, backward smoothing rollouts, and gradient-based parameter updates. Numerical results show that the method recovers the steady-state Riccati solution in a linear-Gaussian case and, in a nonlinear benchmark with heavy-tailed noise, attains accuracy and empirical consistency comparable to the best classical Bayesian and optimization-based estimators and smoothers at a fraction of their online computational cost.
\end{abstract}


\section{Introduction}
\label{sec:introduction}

    State estimation, or filtering, aims to compute online both the filtered estimate of the current state and the one-step-ahead predicted estimate, using the information available up to the current time. Several well-established algorithms address this problem. Bayesian methods include the~\gls{kf}~\cite[Ch.~3.2]{anderson2005optimal} for linear--Gaussian systems, and several methods, such as the~\gls{ekf}~\cite[Ch.~8]{anderson2005optimal}, the~\gls{ukf}~\cite{julier1997new}, and the~\gls{pf}~\cite{arulampalam2002tutorial}, for nonlinear systems. Among optimization-based approaches, \gls{mhe}~\cite{alessandri2008moving,rao2003constrained} is the most established. Complementary to filtering, smoothing aims to reconstruct the complete state trajectory over a finite time interval by exploiting all measurements. In the linear--Gaussian case, the smoothing problem admits an exact solution given by the~\gls{rtss}~\cite{rauch1965maximum}. In nonlinear settings, the problem can be addressed, for instance, through various Bayesian smoothing methods~\cite{sarkka2023bayesian} or by recasting it as a filtering problem on an augmented state-space model~\cite[Ch.~7.2]{anderson2005optimal}. Alternatively, smoothing can be formulated as a finite-horizon optimization problem~\cite{bell1994iterated}. 
    
    The optimization viewpoint is particularly favorable because it provides a unified interpretation of smoothing and recursive state estimation. At a given time instant, smoothing corresponds to solving the~\gls{fie}~\cite{rawlings2020stateestimation} problem over the complete past measurement record, whereas recursive estimation can be cast as a one-step optimization problem in which past information is summarized through the arrival cost~\cite{rawlings2020stateestimation,cox1964estimation}. From a~\gls{dp} perspective, the arrival cost plays the role of a value function~\cite{bertsekas2022abstract}, revealing the common structure underlying~\gls{fie} and optimal control. An exact arrival cost makes this one-step formulation, and more generally any finite-window formulation such as~\gls{mhe}, equivalent to the~\gls{fie}; in practice, however, the arrival cost is generally approximated~\cite{rao2003constrained,robertson1996moving}. This~\gls{dp} viewpoint has recently attracted renewed interest. In particular, Bellman formulations of finite-horizon and recursive estimation problems have been revisited~\cite{moireau2018discrete,lange2024bellman}, clarifying the role of the value function and its relationship with the arrival cost. A distinctive feature of the estimation setting is that the value function evolves \emph{forward} in time as new measurements become available, consistently with the information structure of the problem. In our preliminary work~\cite{menchetti2022linear}, this forward evolution was exploited to learn linear observers through a~\gls{td} approach. 
    
    Beyond the linear case, however, this feature is not exploited by existing learning-based approaches to state estimation inspired by~\gls{rl}~\cite{sutton2018reinforcement}, such as~\cite{morimoto2007reinforcement,cao2021reinforced,straka2026learning}. Indeed, although these methods establish useful connections between estimation and learning, they adopt value-function update mechanisms inherited from optimal control, in which the value function evolves \emph{backward} in time. Consequently, the forward recursion of the estimation value function, which reflects the information structure of the problem, is not exploited in their learning mechanisms.
    
\textbf{Our Contributions.} 
    In this paper, we propose an approach for learning both a recursive state estimator and a smoother from a parametric approximation of the gradient of the optimal value function, which we call the $\bm{\lambda}$-function. Starting from the finite-horizon~\gls{fie} problem, we use~\gls{dp} to derive, under local regularity assumptions, a recursive characterization of the optimal value function. We then show that the $\bm{\lambda}$-function satisfies a forward recursion coupled with the first-order optimality conditions that define the filtered, predicted, and smoothed estimates. This formulation provides the basis for a~\gls{td} learning problem whose targets involve no evaluation of the learned function, since past information enters only through the optimality conditions. This learning problem enables the synthesis of a recursive estimator and a smoother from measurement trajectories alone: unlike supervised learned estimators~\cite{revach2022kalmannet,pratik2021neural,becker2019recurrent}, the proposed method requires no ground-truth states. Measurement-only estimators exist, but either require linear measurement models~\cite{ghosh2024danse} or have only been demonstrated on linear models~\cite{revach2022unsupervised}, while related measurement-only methods target model learning rather than estimation~\cite{krishnan2017structured,simpson2024parallelizable}. Here, instead, the training objective follows from the~\gls{dp} optimality conditions of~\gls{fie} and applies to nonlinear models. Moreover, for an idealized version of the learning problem, we establish local convergence of~\gls{gd} to a critical point under smoothness and regularity assumptions. To make the learning problem computationally tractable, we further develop a practical semi-gradient algorithm, termed $\bm{\lambda}$\emph{-learning}, which alternates between computing estimation and smoothing trajectories for fixed parameters and updating the parameters of the $\bm{\lambda}$-function for fixed trajectories. Numerical experiments show that the method recovers the steady-state Riccati solution in the linear--Gaussian case and, in a nonlinear benchmark with heavy-tailed noise, attains accuracy and empirical consistency comparable to the best classical Bayesian and optimization-based estimators and smoothers at a fraction of their online computational cost.

    The remainder of the paper is organized as follows. Section~\ref{sec:problem_formulation} introduces the~\gls{fie} problem and the recursive state estimation and smoothing problems. Section~\ref{sec:dynamic_programming} develops the~\gls{dp}-based characterization, introduces the $\bm{\lambda}$-function, and characterizes recursive state estimation and smoothing in terms of the $\bm{\lambda}$-function. Section~\ref{sec:learning_the_lambda} formulates the corresponding learning problem as a constrained minimization problem and derives its local reduced reformulation, whose convergence analysis is reported in Appendix~\ref{app:convergence}. Section~\ref{sec:lambda_learning_algorithm} presents the resulting $\bm{\lambda}$\emph{-learning} algorithm as a tractable approximation of the theoretical learning problem. Section~\ref{sec:numerical_experiments} reports numerical results on linear and nonlinear benchmarks, and Section~\ref{sec:conclusions} concludes the paper.
    
\section{Problem Formulation}
\label{sec:problem_formulation}
    
    Consider the following known nonlinear discrete-time state-space model, for time $t=1, 2, \ldots$,
    \begin{equation}
    \label{eq:ss_model}
        \begin{aligned}
            \mathbf{x}_{t+1} &= \mathbf{f}(\mathbf{x}_{t}) + \mathbf{w}_{t},\\
            \mathbf{y}_{t} &= \mathbf{h}(\mathbf{x}_{t}) + \mathbf{v}_{t},
        \end{aligned}
    \end{equation}
    where ${\mathbf{x}_{t} \in \mathbb{R}^{n_{x}}}$ and ${\mathbf{y}_{t} \in \mathbb{R}^{n_{y}}}$ denote the true state and measured output, respectively. The functions $\mathbf{f}$ and $\mathbf{h}$ are the process and measurement functions. The terms $\mathbf{w}_{t} \in \mathbb{R}^{n_{x}}$ and $\mathbf{v}_{t} \in \mathbb{R}^{n_{y}}$ are the process and measurement noise, respectively. To formalize the smoothing and recursive state estimation problems, we adopt an optimization-based formulation. Given a measurement sequence $\mathbf{y}_{1:t}$\footnote{The compact notation $\mathbf{a}_{t_{i}:t_{f}} = \{\mathbf{a}_{t_{i}}, \ldots, \mathbf{a}_{t_{f}}\}$ is used throughout the paper for finite sequences.} available up to time $t$, the unknown state trajectory is estimated by solving  
    \begin{equation}
    \label{eq:opt_ests_t}
        \hat{\mathbf{x}}_{1:t+1|t} \in \arg \min_{\mathbf{s}_{1:t+1}} J_{t}(\mathbf{s}_{1:t}, \mathbf{s}_{t+1}),
    \end{equation}
    where
    \begin{equation*}
        J_{t}(\mathbf{s}_{1:t}, \mathbf{s}_{t+1}) := \sum_{i=1}^{t} \ell_{i}(\mathbf{s}_{i+1}, \mathbf{s}_{i}) + \ell_{0}(\mathbf{s}_{1}).
    \end{equation*}
    Here, $\hat{\mathbf{x}}_{1:t+1|t}$ denotes the sequence of optimal state estimates up to time $t+1$ based on the currently available information, while $\mathbf{s}_{1:t}$ is a candidate state trajectory, with $\mathbf{s}_{i} \in \mathbb{R}^{n_x}$. The terminal variable $\mathbf{s}_{t+1}$ is included in $J_t$ through the last stage cost $\ell_t(\mathbf{s}_{t+1},\mathbf{s}_t)$ and represents the one-step-ahead predicted state decision variable associated with the trajectory $\mathbf{s}_{1:t}$. The dependence of the costs on the realized measurements $\mathbf{y}_{1:t}$ is left implicit, since these measurements are fixed once the information is given. 
    
    Throughout the paper, we distinguish between three state-related quantities: at time $i$, $\mathbf{x}_{i}$ denotes the unknown true state, $\mathbf{s}_{i}$ denotes an optimization variable used to represent the state in the~\gls{fie} problem and in the optimization problems derived below, and, finally, $\hat{\mathbf{x}}_{i|t}$ denotes the optimal estimate of the state obtained using measurements available up to time $t$. Thus, $\hat{\mathbf{x}}_{1:t+1|t}$ is the optimizer associated with $\mathbf{s}_{1:t+1}$, and $\mathbf{x}_{1:t+1}$ is the unknown state trajectory to be reconstructed. Moreover, owing to the additive noise in~\eqref{eq:ss_model}, the noise realizations associated with a candidate trajectory are uniquely given by $\mathbf{s}_{i+1}-\mathbf{f}(\mathbf{s}_i)$ and $\mathbf{y}_i-\mathbf{h}(\mathbf{s}_i)$, so no separate noise decision variables are needed and each stage cost depends only on $(\mathbf{s}_{i+1},\mathbf{s}_i)$. In the following, we do not assume a specific form for the stage and initial costs; they are only required to satisfy the following assumption.
    \begin{assumption}
    \label{ass:twice_differentiability_cost}
        For each $i \geq 1$, the stage cost $\ell_i$ and the initial cost $\ell_0$ are twice continuously differentiable with respect to their arguments.
    \end{assumption}
    A standard choice is to use quadratic costs, i.e.,
    \begin{align}
        \label{eq:stage_cost}
        \ell_{i}(\mathbf{s}_{i+1}, \mathbf{s}_{i}) &:= \frac{1}{2} \|\mathbf{s}_{i+1} - \mathbf{f}(\mathbf{s}_{i})\|^{2}_{\mathbf{Q}^{-1}} + \frac{1}{2} \|\mathbf{y}_{i} - \mathbf{h}(\mathbf{s}_{i})\|^{2}_{\mathbf{R}^{-1}}, \\
        \label{eq:initial_time_cost}
        \ell_{0}(\mathbf{s}_{1}) &:= \frac{1}{2} \|\mathbf{s}_{1} - \mathbf{m}_{1} \|^{2}_{\mathbf{P}_{1}^{-1}}.
    \end{align}
    Here, $\mathbf{Q}\in\mathbb{R}^{n_x\times n_x}$,
    $\mathbf{R}\in\mathbb{R}^{n_y\times n_y}$, and
    $\mathbf{P}_1\in\mathbb{R}^{n_x\times n_x}$ are symmetric positive-definite weighting matrices, which are not required to coincide with noise covariances, since no distributional assumption is made on $\mathbf{w}_t$ and $\mathbf{v}_t$; $\mathbf{m}_1\in\mathbb{R}^{n_x}$ denotes the prior state estimate at the initial time. If $\mathbf{f}$ and $\mathbf{h}$ are twice continuously differentiable, then these costs satisfy Assumption~\ref{ass:twice_differentiability_cost}.
    
    Problem~\eqref{eq:opt_ests_t} encompasses both estimation tasks of interest:
    \begin{itemize}
        \item The \emph{Recursive State Estimation} problem, namely the computation of the filtered and predicted estimates $\hat{\mathbf{x}}_{t|t}$ and $\hat{\mathbf{x}}_{t+1|t}$, which correspond to the last two components of $\hat{\mathbf{x}}_{1:t+1|t}$, from the current measurement $\mathbf{y}_{t}$ and a summary of past information.
        \item The \emph{Smoothing} problem, namely the computation of the complete trajectory $\hat{\mathbf{x}}_{1:T+1|T}$ over a fixed horizon $T$, using the measurements $\mathbf{y}_{1:T}$. This corresponds to solving the~\gls{fie} problem for $t=T$.
    \end{itemize}
    The inclusion of the terminal prediction variable $\hat{\mathbf{x}}_{T+1|T}$ is nonstandard for smoothing, where the trajectory is usually reconstructed only up to time $T$. However, this prediction term allows recursive estimation and smoothing to be derived from the same optimization problem and from the same~\gls{dp}-based interpretation developed in the next section. 

\section{Dynamic Programming Formulation of the FIE Problem}
\label{sec:dynamic_programming}
    
    To define our learning approach, we first derive a~\gls{dp}-based formulation of Problem~\eqref{eq:opt_ests_t}. We identify the optimal value function associated with Problem~\eqref{eq:opt_ests_t}, derive the corresponding Bellman recursion, and characterize recursive state estimation and smoothing in terms of this value function. Its gradient, the $\bm{\lambda}$-function, and the corresponding recursion are then introduced and used to construct the learning formulation. We start by defining the optimal value function at a generic time $t \in \{1, \ldots, T\}$.
    \begin{definition}
    \label{def:value_function}
        The optimal value function is
        \begin{equation}
        \label{eq:V_minimum_J}
            V_t(\mathbf{s}_{t+1}) := \min_{\mathbf{s}_{1:t}} J_t(\mathbf{s}_{1:t}, \mathbf{s}_{t+1}),
        \end{equation}
        where the minimum is assumed to be attained, and
        \begin{equation}
        \label{eq:s_minimizer_V}
            \mathbf{s}_{1:t}^{*}(\mathbf{s}_{t+1}) \in \arg \min_{\mathbf{s}_{1:t}} J_t(\mathbf{s}_{1:t}, \mathbf{s}_{t+1})
        \end{equation}
        is an optimal past trajectory associated with the prescribed $\mathbf{s}_{t+1}$.
    \end{definition}
    The function $V_t$ plays the role of the~\gls{fie} problem's arrival cost~\cite{rawlings2020stateestimation}, summarizes the contribution of past measurements as a function of the terminal variable $\mathbf{s}_{t+1}$, and satisfies the Bellman recursion.
    \begin{proposition}
    \label{prop:Bellman_rec}
        For each $t$ and for all $\mathbf{s}_{t+1} \in \mathbb{R}^{n_{x}}$, if initialized as $V_0(\mathbf{s}_1):=\ell_0(\mathbf{s}_1)$, the optimal value function satisfies the optimal Bellman recursion
        \begin{equation}
            \label{eq:opt_Bellman}
            V_t(\mathbf{s}_{t+1}) = \min_{\mathbf{s}_t} \left(\ell_t(\mathbf{s}_{t+1},\mathbf{s}_t) + V_{t-1}(\mathbf{s}_t)\right). 
        \end{equation}
    \end{proposition}
    \begin{proof}
        The result follows by writing 
        \begin{equation} 
            \label{eq:J_additive_decomposition} 
            J_t(\mathbf{s}_{1:t}, \mathbf{s}_{t+1}) = \ell_t(\mathbf{s}_{t+1},\mathbf{s}_t) + J_{t-1}(\mathbf{s}_{1:t-1}, \mathbf{s}_{t}),
        \end{equation} 
        and minimizing first over $\mathbf{s}_{1:t-1}$ and then over $\mathbf{s}_t$, with $V_0(\mathbf{s}_1) = J_0(\mathbf{s}_1) := \ell_0(\mathbf{s}_1)$. 
    \end{proof}

    In general, multiple minimizers of~\eqref{eq:s_minimizer_V} may be associated with the same terminal value $\mathbf{s}_{t+1}$. While this does not affect the Bellman recursion, the construction of the $\bm{\lambda}$-function requires local uniqueness and smoothness.
    \begin{assumption}
    \label{ass:inner_uniqueness}
        For each $t$ and for every predicted estimate $\hat{\mathbf{x}}_{t+1|t}$ associated with a solution of~\eqref{eq:opt_ests_t}, there exists an open connected neighborhood $\mathcal{X}_{t+1}$ of $\hat{\mathbf{x}}_{t+1|t}$ such that $J_t$ is level-bounded in $\mathbf{s}_{1:t}$ locally uniformly in $\mathbf{s}_{t+1}\in\mathcal{X}_{t+1}$ and, for every $\mathbf{s}_{t+1} \in \mathcal{X}_{t+1}$, the global minimizer $\mathbf{s}^*_{1:t}(\mathbf{s}_{t+1})$ in~\eqref{eq:s_minimizer_V} is unique and satisfies $\nabla^2_{\mathbf{s}_{1:t}} J_t(\mathbf{s}^*_{1:t}(\mathbf{s}_{t+1}), \mathbf{s}_{t+1}) \succ \mathbf{0}$.
    \end{assumption}
    Level-boundedness locally uniformly in $\mathbf{s}_{t+1}$ means that the sublevel sets of $J_t(\cdot,\mathbf{s}_{t+1})$ remain in a common bounded set as $\mathbf{s}_{t+1}$ varies near any point of $\mathcal{X}_{t+1}$~\cite[Def.~1.16]{rockafellar1998variational}. This condition prevents minimizers from escaping to infinity as $\mathbf{s}_{t+1}$ varies; for the quadratic costs~\eqref{eq:stage_cost}--\eqref{eq:initial_time_cost} with continuous $\mathbf{f}$, it holds for any measurement sequence.
    \begin{lemma}
    \label{lemma:value_function_twice_diff}
        Under Assumption~\ref{ass:inner_uniqueness}, for each $t$, the optimal value function $V_t$ is twice continuously differentiable on $\mathcal{X}_{t+1}$.
    \end{lemma}
    \begin{proof}
        By Assumption~\ref{ass:inner_uniqueness} and the parametric minimization theorem~\cite[Thm.~1.17]{rockafellar1998variational}, the unique minimizer $\mathbf{s}_{1:t}^{*}(\mathbf{s}_{t+1})$ depends continuously on $\mathbf{s}_{t+1}\in\mathcal{X}_{t+1}$. Moreover, $\nabla_{\mathbf{s}_{1:t}} J_t(\mathbf{s}_{1:t}^{*}(\mathbf{s}_{t+1}), \mathbf{s}_{t+1}) = \mathbf{0}$, $J_t$ is twice continuously differentiable by Assumption~\ref{ass:twice_differentiability_cost}, and its Hessian with respect to $\mathbf{s}_{1:t}$ is nonsingular by Assumption~\ref{ass:inner_uniqueness}. Hence, by the implicit function theorem~\cite{bertsekas1999nonlinear}, around each $\bar{\mathbf{s}}_{t+1}\in\mathcal{X}_{t+1}$ the stationary point of $J_t(\cdot,\mathbf{s}_{t+1})$ near $\mathbf{s}_{1:t}^{*}(\bar{\mathbf{s}}_{t+1})$ is unique and depends continuously differentiably on $\mathbf{s}_{t+1}$. Being continuous and stationary, $\mathbf{s}_{1:t}^{*}$ coincides with this point, and is thus continuously differentiable on $\mathcal{X}_{t+1}$. So is $V_t(\mathbf{s}_{t+1}) = J_{t}(\mathbf{s}_{1:t}^{*}(\mathbf{s}_{t+1}), \mathbf{s}_{t+1})$, whose gradient, by the chain rule and the first-order optimality condition, is
        \begin{equation}
        \label{eq:grad_V_envelope}
            \nabla_{\mathbf{s}_{t+1}}V_t(\mathbf{s}_{t+1}) = \nabla_{\mathbf{s}_{t+1}} J_t(\mathbf{s}_{1:t}^{*}(\mathbf{s}_{t+1}),
            \mathbf{s}_{t+1}).
        \end{equation}
        Since $J_t$ is twice continuously differentiable and $\mathbf{s}_{1:t}^{*}$ is continuously differentiable, the right-hand side is continuously differentiable, and $V_t$ is twice continuously differentiable on $\mathcal{X}_{t+1}$.
    \end{proof}

    Since $\hat{\mathbf{x}}_{1:t+1|t}$ solves~\eqref{eq:opt_ests_t}, the predicted estimate $\hat{\mathbf{x}}_{t+1|t}$ is a minimizer of $V_t$, and hence a critical point of $V_t$ on $\mathcal{X}_{t+1}$. The next assumption requires it to be the only one.
    \begin{assumption}
    \label{ass:outer_ssosc}
        For each $t$ and for every predicted estimate
        $\hat{\mathbf{x}}_{t+1|t}$ associated with a solution
        of~\eqref{eq:opt_ests_t}, $\hat{\mathbf{x}}_{t+1|t}$ is the unique critical point of $V_t$ on the neighborhood $\mathcal{X}_{t+1}$ of Assumption~\ref{ass:inner_uniqueness}, and hence
        \begin{equation}
        \label{eq:outer_minimization}
            \hat{\mathbf{x}}_{t+1|t} = \arg\min_{\mathbf{s}_{t+1} \in
            \mathcal{X}_{t+1}} V_t(\mathbf{s}_{t+1}).
        \end{equation}
    \end{assumption}
    Under Assumptions~\ref{ass:inner_uniqueness}--\ref{ass:outer_ssosc}, the predicted estimate is therefore the only point of $\mathcal{X}_{t+1}$ where the gradient of $V_t$ vanishes, and the filtered estimate and the smoothing recursion can also be expressed in terms of the optimal value function.
    \begin{corollary}
    \label{cor:value_pred}
        Under Assumption~\ref{ass:inner_uniqueness}, for each $t$ and every $\mathbf{s}_{t+1} \in \mathcal{X}_{t+1}$, the last component of the optimal past trajectory satisfies the smoothing recursion
        \begin{equation}
        \label{eq:smooth_rec_from_V}
            \mathbf{s}^{*}_{t}(\mathbf{s}_{t+1}) = \arg \min_{\mathbf{s}_{t}}
            \left( \ell_{t}(\mathbf{s}_{t+1},\mathbf{s}_{t}) +
            V_{t-1}(\mathbf{s}_{t}) \right).
        \end{equation}
        In particular, $\hat{\mathbf{x}}_{t|t} = \mathbf{s}^{*}_{t}(\hat{\mathbf{x}}_{t+1|t})$.
    \end{corollary}
    \begin{proof}
        Fix $\mathbf{s}_{t+1}\in\mathcal{X}_{t+1}$ and let $\mathbf{s}_t$ be a minimizer in~\eqref{eq:opt_Bellman}. By~\eqref{eq:J_additive_decomposition}, the trajectory $(\mathbf{s}^{*}_{1:t-1}(\mathbf{s}_t),\mathbf{s}_t)$, with $\mathbf{s}^{*}_{1:t-1}(\mathbf{s}_t)$ any optimal past associated with $\mathbf{s}_t$, attains $V_t(\mathbf{s}_{t+1})$ and therefore minimizes $J_t(\cdot,\mathbf{s}_{t+1})$. By Assumption~\ref{ass:inner_uniqueness}, it coincides with $\mathbf{s}^{*}_{1:t}(\mathbf{s}_{t+1})$. Hence the minimizer in~\eqref{eq:opt_Bellman} is unique and equals $\mathbf{s}^{*}_{t}(\mathbf{s}_{t+1})$, which proves~\eqref{eq:smooth_rec_from_V}. Finally, since $\hat{\mathbf{x}}_{1:t+1|t}$ solves~\eqref{eq:opt_ests_t}, its first $t$ components minimize $J_t(\cdot,\hat{\mathbf{x}}_{t+1|t})$, with $\hat{\mathbf{x}}_{t+1|t}\in\mathcal{X}_{t+1}$; by uniqueness, they equal $\mathbf{s}^{*}_{1:t}(\hat{\mathbf{x}}_{t+1|t})$, whose last component is $\hat{\mathbf{x}}_{t|t}$.
    \end{proof}
    
    Corollary~\ref{cor:value_pred} characterizes recursive state estimation and smoothing through the value function. Since $V_t$ is twice continuously differentiable on $\mathcal{X}_{t+1}$ by Lemma~\ref{lemma:value_function_twice_diff}, the same characterization can be expressed through its gradient, which we call the $\bm{\lambda}$-function:
    \begin{equation}
    \label{eq:lambda_def}
        \bm{\lambda}_t(\mathbf{s}_{t+1}) := \nabla_{\mathbf{s}_{t+1}} V_t(\mathbf{s}_{t+1}), \qquad \mathbf{s}_{t+1}\in\mathcal{X}_{t+1}.
    \end{equation}
    Since $\bm{\lambda}_{t-1}$ is defined only on $\mathcal{X}_t$, propagating the $\bm{\lambda}$-function recursively requires the optimal previous state associated with each $\mathbf{s}_{t+1}\in\mathcal{X}_{t+1}$ to remain in $\mathcal{X}_t$.
    \begin{assumption}
    \label{ass:compatible_neighborhoods}
        For each $t$ and every $\mathbf{s}_{t+1} \in \mathcal{X}_{t+1}$, the last component of the optimal past trajectory satisfies $\mathbf{s}_t^*(\mathbf{s}_{t+1}) \in \mathcal{X}_t$, where $\mathcal{X}_t$ is the neighborhood in Assumption~\ref{ass:inner_uniqueness} associated with some predicted estimate $\hat{\mathbf{x}}_{t|t-1}$, and $\mathcal{X}_1 := \mathbb{R}^{n_x}$.
    \end{assumption}
    
    \begin{remark}[\textbf{Local regularity assumptions}]
        Together, Assumptions~\ref{ass:inner_uniqueness}--\ref{ass:compatible_neighborhoods} are local regularity conditions: Assumption~\ref{ass:inner_uniqueness} makes $V_t$ smooth, Assumption~\ref{ass:outer_ssosc} ensures that the first-order condition identifies the prediction, and Assumption~\ref{ass:compatible_neighborhoods} keeps the recursion within the region where $\bm{\lambda}_{t-1}$ is defined. They are stated for every solution of~\eqref{eq:opt_ests_t}, so multiple isolated solutions are allowed. In the linear case with quadratic costs~\eqref{eq:stage_cost}--\eqref{eq:initial_time_cost}, they hold globally with $\mathcal{X}_{t+1}=\mathbb{R}^{n_x}$, since $V_t$ is a strictly convex quadratic function with Hessian $\mathbf{P}_{t+1|t}^{-1}$, where $\mathbf{P}_{t+1|t}$ is the predicted covariance of the~\gls{kf} with weighting matrices $\mathbf{Q}$, $\mathbf{R}$, and $\mathbf{P}_1$~\cite{menchetti2022linear}.
    \end{remark}
    
    \begin{theorem}
    \label{theo:ext_lambda}
        Under Assumptions~\ref{ass:inner_uniqueness}
        and~\ref{ass:compatible_neighborhoods}, for each $t$ and every
        $\mathbf{s}_{t+1}\in\mathcal{X}_{t+1}$, with
        $\bm{\lambda}_{0}(\mathbf{s}_{1}) := \nabla_{\mathbf{s}_{1}}
        \ell_{0}(\mathbf{s}_{1})$,
        \begin{align}
        \label{eq:lambda_from_ell}
            \bm{\lambda}_t(\mathbf{s}_{t+1}) &= \nabla_{\mathbf{s}_{t+1}}
            \ell_{t}(\mathbf{s}_{t+1}, \mathbf{s}_{t}^{*}), \\
        \label{eq:s_from_lambda}
            \mathbf{0} &= \nabla_{\mathbf{s}_{t}} \ell_{t}(\mathbf{s}_{t+1},
            \mathbf{s}_{t}^{*}) + \bm{\lambda}_{t-1}(\mathbf{s}_{t}^{*}),
        \end{align}
        where $\mathbf{s}_{t}^{*}:=\mathbf{s}_{t}^{*}(\mathbf{s}_{t+1})$.
    \end{theorem}
    \begin{proof}
        By~\eqref{eq:grad_V_envelope}, $\bm{\lambda}_t(\mathbf{s}_{t+1})$ equals the partial gradient of $J_t$ with respect to $\mathbf{s}_{t+1}$ at $(\mathbf{s}^*_{1:t}(\mathbf{s}_{t+1}), \mathbf{s}_{t+1})$. Since $\ell_t$ is the only term of $J_t$ that depends on $\mathbf{s}_{t+1}$, this proves~\eqref{eq:lambda_from_ell}. By Corollary~\ref{cor:value_pred}, $\mathbf{s}_t^*$ minimizes $\ell_t(\mathbf{s}_{t+1},\cdot) + V_{t-1}(\cdot)$. By Assumption~\ref{ass:compatible_neighborhoods}, $\mathbf{s}_t^*\in\mathcal{X}_t$, where $V_{t-1}$ is differentiable with gradient $\bm{\lambda}_{t-1}$ by Lemma~\ref{lemma:value_function_twice_diff} (for $t=1$, by $V_0=\ell_0$). The first-order optimality condition at $\mathbf{s}_t^*$ then yields~\eqref{eq:s_from_lambda}.
    \end{proof}
    Theorem~\ref{theo:ext_lambda} shows that the $\bm{\lambda}$-function satisfies its own recursion and can be propagated without evaluating the value function. In particular, $\bm{\lambda}_t$ is given by the partial gradient of the current stage cost, while $\bm{\lambda}_{t-1}$ enters only through the optimality condition~\eqref{eq:s_from_lambda} that determines $\mathbf{s}_t^*(\mathbf{s}_{t+1})$. This structure will yield a learning problem whose targets involve no evaluation of the learned function.

     \begin{remark}[\textbf{Relation with value-gradient methods in control}]
    \label{rem:control_lambda}
        The $\bm{\lambda}$-function is the estimation counterpart of the costate function learned by value-gradient methods in approximate dynamic programming, such as dual heuristic programming and its globalized variant~\cite{prokhorov1997adaptive,kim2020convergence}, and value-gradient learning~\cite{fairbank2012value}. For a control problem with dynamics $\mathbf{s}_{t+1}=\mathbf{g}(\mathbf{s}_t,\mathbf{a}_t)$, stage cost $c_t(\mathbf{s}_t,\mathbf{a}_t)$, and optimal action $\mathbf{a}_t^*$, the costate $\bm{\mu}_t:=\nabla V_t$ satisfies
        \begin{equation*}
            \bm{\mu}_t(\mathbf{s}_t) = \nabla_{\mathbf{s}_t} c_t(\mathbf{s}_t,\mathbf{a}_t^*) + \nabla_{\mathbf{s}_t}\mathbf{g}(\mathbf{s}_t,\mathbf{a}_t^*)^{\top} \bm{\mu}_{t+1}\big(\mathbf{g}(\mathbf{s}_t,\mathbf{a}_t^*)\big),
        \end{equation*}
        which propagates backward in time and whose targets explicitly evaluate the costate at the next time step. This contrasts with the forward recursion~\eqref{eq:lambda_from_ell}--\eqref{eq:s_from_lambda}, in which $\bm{\lambda}_{t-1}$ enters only implicitly, through the optimality condition.
    \end{remark}
     
    Furthermore, the characterizations in Corollary~\ref{cor:value_pred} can also be rewritten in terms of the $\bm{\lambda}$-function alone.
    \begin{corollary}
    \label{cor:lambda_opt}
        Under Assumptions~\ref{ass:inner_uniqueness}--\ref{ass:compatible_neighborhoods}, let $\bar{\mathbf{s}}_{t+1}\in\mathcal{X}_{t+1}$ and $\bar{\mathbf{s}}_{t}\in\mathcal{X}_{t}$ be arbitrary reference points. The predicted estimate is locally characterized in terms of the $\bm{\lambda}$-function by
        \begin{equation}
        \label{eq:pred_from_lambda_integral}
            \hat{\mathbf{x}}_{t+1|t} = \arg\min_{\mathbf{s}_{t+1}\in\mathcal{X}_{t+1}} \int_{\Omega_{t+1}} \bm{\lambda}_t(\mathbf{z})^{\top}\operatorname{d}\mathbf{z},
        \end{equation}
        where $\Omega_{t+1}$ is any continuously differentiable path contained in $\mathcal{X}_{t+1}$ joining $\bar{\mathbf{s}}_{t+1}$ to $\mathbf{s}_{t+1}$. The corresponding first-order optimality condition, which by Assumption~\ref{ass:outer_ssosc} holds on $\mathcal{X}_{t+1}$ only at $\hat{\mathbf{x}}_{t+1|t}$, is
        \begin{equation}
        \label{eq:lambda_pred_zero}
            \nabla_{\mathbf{s}_{t+1}} \ell_t(\hat{\mathbf{x}}_{t+1|t}, \hat{\mathbf{x}}_{t|t}) = \mathbf{0}.
        \end{equation}
        Moreover, for every $\mathbf{s}_{t+1}\in\mathcal{X}_{t+1}$, the last component of the associated optimal past trajectory satisfies the recursion
        \begin{equation}
        \label{eq:smooth_rec_from_lambda_integral}
            \mathbf{s}^{*}_{t}(\mathbf{s}_{t+1}) = \arg\min_{\mathbf{s}_{t} \in \mathcal{X}_{t}} \left[ \ell_t(\mathbf{s}_{t+1},\mathbf{s}_{t}) + \int_{\Omega_{t}} \bm{\lambda}_{t-1}(\mathbf{z})^{\top}
            \operatorname{d}\mathbf{z} \right],
        \end{equation}
        where $\Omega_{t}$ is any continuously differentiable path contained in $\mathcal{X}_{t}$ joining $\bar{\mathbf{s}}_{t}$ to $\mathbf{s}_{t}$. The corresponding first-order optimality condition is~\eqref{eq:s_from_lambda}. In particular, evaluating \eqref{eq:smooth_rec_from_lambda_integral} and~\eqref{eq:s_from_lambda} at $\mathbf{s}_{t+1}=\hat{\mathbf{x}}_{t+1|t}$ gives $\hat{\mathbf{x}}_{t|t} = \mathbf{s}^{*}_{t}(\hat{\mathbf{x}}_{t+1|t})$ and
        \begin{equation}
        \label{eq:lambda_upd_foc}
            \nabla_{\mathbf{s}_{t}} \ell_t(\hat{\mathbf{x}}_{t+1|t}, \hat{\mathbf{x}}_{t|t}) + \bm{\lambda}_{t-1}(\hat{\mathbf{x}}_{t|t}) = \mathbf{0}.
        \end{equation}
    \end{corollary}
    \begin{proof}
        Since $\bm{\lambda}_t=\nabla V_t$ on the open connected set $\mathcal{X}_{t+1}$, for every $\mathbf{s}_{t+1}\in\mathcal{X}_{t+1}$,
        \begin{equation*}
            V_t(\mathbf{s}_{t+1}) = V_t(\bar{\mathbf{s}}_{t+1}) + \int_{\Omega_{t+1}} \bm{\lambda}_t(\mathbf{z})^{\top} \operatorname{d}\mathbf{z},
        \end{equation*}
        independently of the path, and similarly for $V_{t-1}$ on $\mathcal{X}_t$. Since the constants do not affect the minimizers, \eqref{eq:pred_from_lambda_integral} is equivalent to~\eqref{eq:outer_minimization}. Moreover, by Assumption~\ref{ass:compatible_neighborhoods}, the minimizer in~\eqref{eq:smooth_rec_from_V} lies in $\mathcal{X}_t$, so restricting the minimization to $\mathcal{X}_t$ does not change it, and~\eqref{eq:smooth_rec_from_lambda_integral} is equivalent to~\eqref{eq:smooth_rec_from_V}. The optimality conditions follow from~\eqref{eq:lambda_from_ell}, \eqref{eq:s_from_lambda}, and $\bm{\lambda}_t(\hat{\mathbf{x}}_{t+1|t})=\mathbf{0}$.
    \end{proof}
    This corollary shows that the $\bm{\lambda}$-function carries sufficient information to characterize both smoothing and recursive state estimation, mirroring Corollary~\ref{cor:value_pred}. In particular, the integral representation shows the equivalence with the value-function formulation, while the optimality conditions~\eqref{eq:s_from_lambda},~\eqref{eq:lambda_pred_zero}, and~\eqref{eq:lambda_upd_foc} provide the relations that will be used in the learning formulation. Moreover, in this characterization, the predicted estimate is identified first, as the unique zero of $\bm{\lambda}_t$ on $\mathcal{X}_{t+1}$, and the filtered estimate is then obtained by applying the smoothing recursion to it, $\hat{\mathbf{x}}_{t|t}=\mathbf{s}_t^*(\hat{\mathbf{x}}_{t+1|t})$. Hence, filtering and smoothing are defined by the same map, and the order is reversed with respect to classical recursive filters, in which the predicted estimate is obtained by propagating the filtered estimate through the process model. The last results provide the basis for the parametrized learning formulation developed in the next section. The learning problem is constructed by enforcing the recursion~\eqref{eq:lambda_from_ell} together with the optimality conditions~\eqref{eq:s_from_lambda},~\eqref{eq:lambda_pred_zero}, and~\eqref{eq:lambda_upd_foc}. 

\section{Learning the \texorpdfstring{$\bm{\lambda}$}{lambda}-Function}
\label{sec:learning_the_lambda}

    The previous section showed that $\bm{\lambda}_t$ can be used to characterize recursive state estimation and smoothing. This suggests that~\gls{td} methods~\cite[Ch.~6]{sutton2018reinforcement} are well suited for learning the $\bm{\lambda}$-function by exploiting the forward recursion~\eqref{eq:lambda_from_ell}, together with the optimality conditions defining smoothing, prediction, and filtering, namely~\eqref{eq:s_from_lambda}, \eqref{eq:lambda_pred_zero}, and~\eqref{eq:lambda_upd_foc}. This leads to a~\gls{td}-based learning formulation defined directly on the $\bm{\lambda}$-function, in which the learning residual measures the mismatch between the parametric approximation of $\bm{\lambda}_t$ and the gradient of the current stage cost evaluated along trajectories satisfying the recursive optimality conditions. Unlike value-function~\gls{td} learning, the resulting targets involve no evaluation of the learned function, since past information enters only through the optimality conditions that generate the trajectories. This removes one known source of instability of bootstrapped value-function targets under function approximation~\cite{baird1995residual,tsitsiklis1996analysis}.
    
\subsection{Optimization-Based Learning Formulation}

    We formulate the learning problem by imposing the optimality conditions defining smoothing, prediction, and filtering as equality constraints, while the forward recursion for the $\bm{\lambda}$-function is treated as the~\gls{td} error to be minimized in squared norm. Let $\mathcal{D}:=\{\mathbf{Y},\mathbf{M}_{1}\}$ denote the dataset, where $\mathbf{Y}\in\mathbb{R}^{N\,T\,n_y}$ stacks all measurements from $N$ trajectories of length $T$. Specifically, the measurements are stacked by time as
    \begin{equation*}
        \mathbf{Y} :=
        \begin{bmatrix}
            \mathbf{Y}_{1}^{\top} & \cdots & \mathbf{Y}_{T}^{\top}
        \end{bmatrix}^{\top}, \,
        \mathbf{Y}_{t}:=
        \begin{bmatrix}
            (\mathbf{y}_{t}^{(1)})^{\top}
            & \cdots &
            (\mathbf{y}_{t}^{(N)})^{\top}
        \end{bmatrix}^{\top}, 
    \end{equation*}
    where $\mathbf{y}_{t}^{(k)}$ is the measurement at time $t$ for trajectory $k$. The vector $\mathbf{M}_{1}\in\mathbb{R}^{Nn_x}$ is obtained by repeating the initial prior vector $\mathbf{m}_{1}$ across all trajectories, so as to preserve the same batch structure as the general measurement blocks $\mathbf{Y}_{t}$. The dataset therefore contains measurements and initial prior information, but \emph{no ground-truth state trajectories}. With the same time-wise structure as $\mathbf{Y}$, the smoothed-state decision vector $\mathbf{S} \in \mathbb{R}^{N \, T \, n_{x}}$ is defined as 
    \begin{equation*}
        \mathbf{S} := 
        \begin{bmatrix}
            \mathbf{S}_{1}^{\top} & \cdots & \mathbf{S}_{T}^{\top} 
        \end{bmatrix}^{\top}, \,
        \mathbf{S}_{t} :=
        \begin{bmatrix}
            (\mathbf{s}_{t}^{(1)})^{\top} & \cdots & (\mathbf{s}_{t}^{(N)})^{\top} 
        \end{bmatrix}^{\top},
    \end{equation*}
    where $\mathbf{s}_{t}^{(k)}$ is the state at time $t$ of the candidate smoothed trajectory associated with trajectory $k$. Similarly, let $\mathbf{X}\in\mathbb{R}^{2NTn_x}$ be the candidate predicted and filtered estimates as
    \begin{equation*}
        \mathbf{X} :=
        \begin{bmatrix}
            (\mathbf{X}_1^p)^\top & (\mathbf{X}_1^u)^\top & \cdots &
            (\mathbf{X}_T^p)^\top & (\mathbf{X}_T^u)^\top
        \end{bmatrix}^\top,
    \end{equation*}
    where $\mathbf{X}_t^p,\mathbf{X}_t^u\in\mathbb{R}^{N\,n_x}$ are defined by
    \begin{align*}
        \mathbf{X}_{t}^{p} &:= 
        \begin{bmatrix}
            (\mathbf{p}_{t}^{(1)})^{\top} & \cdots & (\mathbf{p}_{t}^{(N)})^{\top}
        \end{bmatrix}^{\top}, \\
        \mathbf{X}_{t}^{u} &:= 
        \begin{bmatrix}
            (\mathbf{u}_{t}^{(1)})^{\top} & \cdots & (\mathbf{u}_{t}^{(N)})^{\top}
        \end{bmatrix}^{\top}.
    \end{align*}
    Here, $\mathbf{p}_{t}^{(k)},\mathbf{u}_{t}^{(k)}\in\mathbb{R}^{n_x}$ are additional decision variables representing, respectively, the candidate one-step-ahead predicted estimate and filtered estimate at time $t$ for trajectory $k$. At the solution of the learning problem, the corresponding optimal values of $\mathbf{S}$ and $\mathbf{X}$ are denoted by $\hat{\mathbf{S}}$ and $\hat{\mathbf{X}}$, respectively. In particular, $\hat{\mathbf{S}}$ collects the learned smoothed trajectories $\hat{\mathbf{x}}_{1:T|T}^{(k)}$ while $\hat{\mathbf{X}}$ collects the corresponding learned predicted and filtered estimate trajectories, with components $\hat{\mathbf{x}}_{t+1|t}^{(k)}$ and $\hat{\mathbf{x}}_{t|t}^{(k)}$ at time $t$. With a slight abuse of notation, we use the same symbols for the estimates induced by $\tilde{\bm{\lambda}}$ as for the exact ones in Section~\ref{sec:dynamic_programming}.

    \begin{remark}[\textbf{Choice of the hypothesis space for $\bm{\lambda}(\cdot)$}]
        Since the objective is to construct an approximation of the $\bm{\lambda}$-function suitable for both recursive state estimation and smoothing, we choose a time-invariant parametrization $\tilde{\bm{\lambda}}(\mathbf{s},\mathbf{p};\bm{\theta})$, where $\bm{\theta}\in\mathbb{R}^{n_\theta}$ is the parameter vector. The first argument represents the variable at which the $\bm{\lambda}$-function is evaluated, while the second argument is the corresponding candidate predicted estimate. Including $\mathbf{p}$ as an additional argument is consistent both with the analytical structure obtained in~\cite{menchetti2022linear,lange2024bellman}, where the gradient of the value function depends linearly on $\mathbf{s}-\mathbf{p}$, and with the local characterization in Section~\ref{sec:dynamic_programming}, where the $\bm{\lambda}$-function is defined around the predicted estimate. This choice is an approximation. Indeed, the exact $\bm{\lambda}_t$ is time-varying and depends on the whole measurement history, which $\tilde{\bm{\lambda}}$ summarizes through the predicted estimate. However, in the linear--Gaussian case at steady state, this parametrization is exact. Replacing $\bm{\lambda}_{t-1}$ by $\tilde{\bm{\lambda}}$ in the optimality conditions~\eqref{eq:s_from_lambda}, \eqref{eq:lambda_pred_zero}, and~\eqref{eq:lambda_upd_foc} shows that the candidate predicted and filtered estimates at time $t$ are determined by the current measurement and the previous predicted estimate, and the smoothed estimate additionally by the next smoothed state. This mirrors the recursive structure of classical smoothers, such as the~\gls{rtss}~\cite{rauch1965maximum}, while allowing for nonlinear models. Other parametrizations could in principle be considered, leading to different structural properties.
    \end{remark}

    The relations~\eqref{eq:lambda_from_ell},~\eqref{eq:s_from_lambda},~\eqref{eq:lambda_pred_zero}, and~\eqref{eq:lambda_upd_foc} are no longer automatically satisfied when the \emph{true} ${\bm{\lambda}}$ is replaced by an approximation $\tilde{\bm{\lambda}}$. Hence, to define our learning problem, we introduce the following feasibility assumption.
    \begin{assumption}
    \label{ass:feasibility_optimal}
        For the dataset $\mathcal{D}$ and the parametric class $\tilde{\bm{\lambda}}$, there exists at least one parameter vector $\bm{\theta}^{\circ}\in\mathbb{R}^{n_{\theta}}$ and at least one pair of decision variables $(\mathbf{S}^{\circ},\mathbf{X}^{\circ})$ such that the smoothing, prediction, and filtering optimality conditions induced by $\tilde{\bm{\lambda}}$, corresponding to~\eqref{eq:s_from_lambda},~\eqref{eq:lambda_pred_zero}, and~\eqref{eq:lambda_upd_foc}, are satisfied on the dataset.
    \end{assumption}

    Assumption~\ref{ass:feasibility_optimal} only concerns the feasibility of the optimality conditions associated with smoothing, prediction, and filtering. It does not require the parametric approximation to satisfy the $\bm{\lambda}$-recursion~\eqref{eq:lambda_from_ell} exactly. The violation of this recursion is instead measured by the corresponding~\gls{td} error and minimized in the learning problem below. This distinction reflects the different roles of the two sets of equations: the optimality conditions define the smoothed, predicted and filtered trajectories, induced by a given $\tilde{\bm{\lambda}}$, whereas the recursion~\eqref{eq:lambda_from_ell} is used to provide the~\gls{td} fitting condition to learn the parameters. Under this assumption, the ideal learning problem for the parametrized $\bm{\lambda}$-function is defined as
    \begin{subequations}
        \begin{align}
            \min_{\bm{\theta},\mathbf{S},\mathbf{X}}\quad & \rho \|\bm{\theta}\|_{2}^{2} + \|\bm{\delta}(\mathbf{S}, \mathbf{X}; \bm{\theta})\|^{2}_{2} \\
            \text{s.t.}\quad & \mathbf{r}_{S}(\mathbf{S}, \mathbf{X};\bm{\theta}) = \mathbf{0}, \label{eq:R2eq}\\
            & \mathbf{r}_{X}(\mathbf{X};\bm{\theta}) = \mathbf{0},\label{eq:R3eq} 
        \end{align}
        \label{eq:complete_training_min}
    \end{subequations}
    where we define the residuals
    \begin{equation}
        \label{eq:explicit_delta_theta}
        \bm{\delta}(\mathbf{S}, \mathbf{X}; \bm{\theta}) = 
        \begin{bmatrix}
            \tilde{\bm{\lambda}}(\overline{\mathbf{S}}_{T+1}, \mathbf{X}_{T}^{p}; \bm{\theta}) - \nabla_{\mathbf{S}_{T+1}}\ell_{T}(\overline{\mathbf{S}}_{T+1}, \mathbf{S}_{T}) \\
            \vdots \\
            \tilde{\bm{\lambda}}(\mathbf{S}_{2}, \mathbf{X}_{1}^{p}; \bm{\theta}) - \nabla_{\mathbf{S}_{2}}\ell_{1}(\mathbf{S}_{2}, \mathbf{S}_{1}) \\
            \tilde{\bm{\lambda}}(\mathbf{S}_{1}, \mathbf{M}_{1}; \bm{\theta}) - \nabla_{\mathbf{S}_{1}}\ell_{0}(\mathbf{S}_{1}) 
        \end{bmatrix},
    \end{equation}
    \begin{equation}
        \label{eq:explicit_RS}
        \mathbf{r}_{S}(\mathbf{S}, \mathbf{X};\bm{\theta}) = 
        \begin{bmatrix}
            \tilde{\bm{\lambda}}(\mathbf{S}_{T}, \mathbf{X}_{T-1}^{p}; \bm{\theta}) + \nabla_{\mathbf{S}_{T}}\ell_{T}(\overline{\mathbf{S}}_{T+1}, \mathbf{S}_{T}) \\
            \vdots \\
            \tilde{\bm{\lambda}}(\mathbf{S}_{2}, \mathbf{X}_{1}^{p}; \bm{\theta}) + \nabla_{\mathbf{S}_{2}}\ell_{2}(\mathbf{S}_{3}, \mathbf{S}_{2}) \\
            \tilde{\bm{\lambda}}(\mathbf{S}_{1}, \mathbf{M}_{1}; \bm{\theta}) + \nabla_{\mathbf{S}_{1}}\ell_{1}(\mathbf{S}_{2}, \mathbf{S}_{1}) 
        \end{bmatrix}, 
    \end{equation}
    \begin{equation}
        \label{eq:explicit_RX}
        \mathbf{r}_{X}(\mathbf{X};\bm{\theta}) = 
        \begin{bmatrix}
            \nabla_{\mathbf{X}_{T}^{p}}\ell_{T}(\mathbf{X}^{p}_{T}, \mathbf{X}^{u}_{T}) \\
            \tilde{\bm{\lambda}}(\mathbf{X}^{u}_{T}, \mathbf{X}_{T-1}^{p}; \bm{\theta}) + \nabla_{\mathbf{X}_{T}^{u}}\ell_{T}(\mathbf{X}^{p}_{T}, \mathbf{X}^{u}_{T}) \\
            \vdots \\
            \nabla_{\mathbf{X}_{2}^{p}}\ell_{2}(\mathbf{X}_{2}^{p}, \mathbf{X}_{2}^{u}) \\
            \tilde{\bm{\lambda}}(\mathbf{X}^{u}_{2}, \mathbf{X}_{1}^{p}; \bm{\theta}) + \nabla_{\mathbf{X}_{2}^{u}}\ell_{2}(\mathbf{X}_{2}^{p}, \mathbf{X}_{2}^{u}) \\
            \nabla_{\mathbf{X}_{1}^{p}}\ell_{1}(\mathbf{X}^{p}_{1}, \mathbf{X}^{u}_{1}) \\
            \tilde{\bm{\lambda}}(\mathbf{X}^{u}_{1}, \mathbf{M}_{1}; \bm{\theta}) + \nabla_{\mathbf{X}_{1}^{u}}\ell_{1}(\mathbf{X}^{p}_{1}, \mathbf{X}^{u}_{1})
        \end{bmatrix}.
    \end{equation}
    In the residuals, with a slight abuse of notation, $\tilde{\bm{\lambda}}$, the stage costs $\ell_t$, and the corresponding gradients are evaluated over entire trajectories, rather than for a single sample.

    Problem~\eqref{eq:complete_training_min} can be interpreted as a~\gls{td} fitting problem for the parametrized $\bm{\lambda}$-function, constrained to trajectories that satisfy the corresponding smoothing, prediction, and filtering optimality conditions. The objective promotes regularization of the parameter vector while penalizing the~\gls{td} error $\bm{\delta}$ in a nonlinear least-squares sense, with $\rho>0$ controlling the trade-off between these two terms. More precisely, $\bm{\delta}$ measures the mismatch between the parametrized approximation $\tilde{\bm{\lambda}}$ and the target gradients induced by the $\bm{\lambda}$-recursion~\eqref{eq:lambda_from_ell}. By contrast, $\mathbf{r}_{S}$ and $\mathbf{r}_{X}$ encode the optimality conditions required by the $\bm{\lambda}$-based characterization of smoothing and recursive estimation computed using the parametrized $\bm{\lambda}$-function. Specifically, $\mathbf{r}_{S}$ enforces the smoothing optimality condition~\eqref{eq:s_from_lambda}, while $\mathbf{r}_{X}$ enforces the prediction and filtering optimality conditions~\eqref{eq:lambda_pred_zero}--\eqref{eq:lambda_upd_foc}. Their dimensions match those of the corresponding decision variables, so that, for fixed $\bm{\theta}$ and terminal vector $\overline{\mathbf{S}}_{T+1}$, the constraints are intended to determine the predicted, filtered, and smoothed trajectories. Moreover, their block structure reflects the forward recursion of prediction-filtering and the backward recursion of smoothing.

    \begin{remark} [\textbf{The role of the fixed terminal vector}]
        Both $\bm{\delta}$ and $\mathbf{r}_{S}$ involve the fixed terminal vector
        \begin{equation*}
            \overline{\mathbf{S}}_{T+1} :=
            \begin{bmatrix}
                \left(\overline{\mathbf{s}}_{T+1}^{(1)}\right)^{\top} & \cdots & \left(\overline{\mathbf{s}}_{T+1}^{(N)}\right)^{\top}
            \end{bmatrix}^{\top}
            \in \mathbb{R}^{N n_x},
        \end{equation*}
        which collects the prescribed terminal states used to initialize the backward smoothing recursion. Ideally, $\overline{\mathbf{S}}_{T+1}$ should be given or, alternatively, be set equal to the stacked optimal predicted estimates at time $T+1$, namely $\hat{\mathbf{X}}_{T}^{p}$. However, these estimates are themselves part of the quantities to be reconstructed and are therefore not available a priori. Thus, Problem~\eqref{eq:complete_training_min} should be understood as a learning problem posed for a prescribed terminal initialization $\overline{\mathbf{S}}_{T+1}$. In the practical implementation, this terminal vector is updated iteratively using the predicted estimate obtained from the current forward estimation step, possibly perturbed for exploration. 
    \end{remark}

\subsection{Local Reformulation and Convergence}

    We now derive a local unconstrained reformulation of~\eqref{eq:complete_training_min}, showing that the optimality conditions select smoothed, predicted, and filtered trajectories as implicit functions of $\bm{\theta}$, and that, once these trajectories are substituted into the learning residual, the resulting objective depends only on the $\bm{\lambda}$-based~\gls{td} error $\bm{\delta}$ evaluated along trajectories satisfying the estimation optimality conditions. This yields a smooth local problem in the parameter vector, suitable for standard gradient-based convergence arguments. To characterize the local unconstrained reformulation, we introduce the following assumption.
    \begin{assumption}
    \label{ass:smoothness_and_existence}
        The residuals $\bm{\delta}$, $\mathbf{r}_{S}$, and $\mathbf{r}_{X}$ satisfy the following conditions.
        \begin{itemize}
            \item The residuals $\bm{\delta}$,  $\mathbf{r}_{S}$, and $\mathbf{r}_{X}$ are continuously differentiable. 
            \item The Jacobians $\nabla_{\mathbf{S}}\mathbf{r}_{S}(\mathbf{S},\mathbf{X};\bm{\theta})$ and $\nabla_{\mathbf{X}}\mathbf{r}_{X}(\mathbf{X};\bm{\theta})$ are nonsingular in an open neighborhood $\mathcal{U}$ of the feasible point $(\mathbf{S}^{\circ},\mathbf{X}^{\circ},\bm{\theta}^{\circ})$ of Assumption~\ref{ass:feasibility_optimal}.
        \end{itemize}
    \end{assumption}
    Since $\bm{\delta}$, $\mathbf{r}_{S}$, and $\mathbf{r}_{X}$ are built from the stage cost, the initial cost, and the parametrized approximation $\tilde{\bm{\lambda}}$, and since the costs are twice continuously differentiable, this requirement mainly restricts the choice of the approximating class of $\tilde{\bm{\lambda}}$ to smooth maps, such as neural networks with smooth activations, radial basis expansions, or polynomial models. The nonsingularity of $\nabla_{\mathbf{X}}\mathbf{r}_{X}$ and $\nabla_{\mathbf{S}}\mathbf{r}_{S}$ guarantees that the prediction-filtering constraints locally determine $\mathbf{X}$ as a function of $\bm{\theta}$, while the smoothing constraints locally determine $\mathbf{S}$ as a function of $(\mathbf{X},\bm{\theta})$. This yields the local implicit functions formalized next.
    
    \begin{lemma}[\textbf{Implicit function representation}]
    \label{lem:implicit_representation}
        Let Assumption~\ref{ass:smoothness_and_existence} hold. Then, there exist neighborhoods $\mathcal{U}_{\theta}$ of $\bm{\theta}^{\circ}$, $\mathcal{U}_{X}$ of $\mathbf{X}^{\circ}$, and $\mathcal{U}_{S}$ of $\mathbf{S}^{\circ}$, with $\mathcal{U}_{S} \times \mathcal{U}_{X} \times \mathcal{U}_{\theta} \subseteq \mathcal{U}$, such that the constraint equations $\mathbf{r}_{S}(\mathbf{S},\mathbf{X};\bm{\theta}) = \mathbf{0}$ and $\mathbf{r}_{X}(\mathbf{X};\bm{\theta}) = \mathbf{0}$ locally define two continuously differentiable implicit functions $\hat{\mathbf{X}}(\bm{\theta})$ and $\hat{\mathbf{S}}(\bm{\theta})$. These functions satisfy $\mathbf{r}_{S}(\hat{\mathbf{S}}(\bm{\theta}),\hat{\mathbf{X}}(\bm{\theta});\bm{\theta}) = \mathbf{0}$, $\mathbf{r}_{X}(\hat{\mathbf{X}}(\bm{\theta});\bm{\theta}) = \mathbf{0}$, for all $\bm{\theta}\in\mathcal{U}_{\theta}$. Moreover, the conditions $\hat{\mathbf{X}}(\bm{\theta}^{\circ})=\mathbf{X}^{\circ}$, and $\hat{\mathbf{S}}(\bm{\theta}^{\circ})=\mathbf{S}^{\circ}$ hold.
    \end{lemma}
    \begin{proof}
        The result follows from a sequential application of the implicit function theorem~\cite{bertsekas1999nonlinear}. First, applying the theorem to $\mathbf{r}_{X}(\mathbf{X};\bm{\theta})=\mathbf{0}$ yields a local continuously differentiable implicit function $\hat{\mathbf{X}}(\bm{\theta})$ with $\hat{\mathbf{X}}(\bm{\theta}^{\circ})=\mathbf{X}^{\circ}$. Since $\mathbf{r}_{S}$ and $\hat{\mathbf{X}}$ are continuously differentiable, the composition $\mathbf{r}_{S}(\mathbf{S},\hat{\mathbf{X}}(\bm{\theta});\bm{\theta})$ is continuously differentiable. Applying the theorem again with respect to $\mathbf{S}$ yields a local continuously differentiable implicit function $\hat{\mathbf{S}}(\bm{\theta})$ with $\hat{\mathbf{S}}(\bm{\theta}^{\circ})=\mathbf{S}^{\circ}$.
    \end{proof}
    Lemma~\ref{lem:implicit_representation} justifies the local elimination of $\mathbf{S}$ and $\mathbf{X}$ from~\eqref{eq:complete_training_min}. Hence, for $\bm{\theta}\in\mathcal{U}_{\theta}$, substituting the implicit trajectory maps into~\eqref{eq:complete_training_min} yields, for the fixed terminal vector $\overline{\mathbf{S}}_{T+1}$, the local optimization problem
    \begin{subequations}
    \begin{align}
    \label{eq:reduced_min}
        &\min_{\bm{\theta}\in\mathcal{U}_{\theta}} \Psi(\bm{\theta}) \\
    \text{where} \notag \\
    \label{eq:psi}
        &\Psi(\bm{\theta}) = \left\|\bm{\delta}\big(\hat{\mathbf{S}}(\bm{\theta}),\hat{\mathbf{X}}(\bm{\theta}),\bm{\theta}\big)\right\|^{2}_{2}+ \rho\|\bm{\theta}\|^{2}_{2}.
    \end{align}
    \end{subequations}
    The reduced objective~\eqref{eq:psi} makes the~\gls{td} structure of the proposed formulation explicit, since after the local reconstruction of trajectories satisfying the estimation and smoothing optimality conditions, the parameter vector is learned by minimizing only the regularized $\bm{\lambda}$-based~\gls{td} residual. The formulation is structurally close to standard~\gls{td} learning, but retains the dependence of the reconstructed trajectories on $\bm{\theta}$ through the implicit maps $\hat{\mathbf{X}}(\bm{\theta})$ and $\hat{\mathbf{S}}(\bm{\theta})$. Moreover, the reduced formulation provides a favorable setting for studying gradient-based updates directly in the parameter space. Under suitable local regularity and smoothness conditions, together with a margin condition on the initial level set,~\gls{gd} applied to~\eqref{eq:reduced_min} can be shown to remain within the region where the implicit trajectory maps are well defined, to generate a sequence of finite length, and to converge to a critical point of the reduced objective. Establishing this result is particularly important because the proposed formulation originates from an undiscounted~\gls{td}-based characterization, for which standard fixed-point convergence arguments~\cite{bertsekas2022abstract} are not directly applicable. The reduced formulation therefore makes it possible to replace such arguments with tools from smooth nonconvex optimization and to establish convergence of~\gls{gd} on the reduced objective. The complete convergence analysis is given in Appendix~\ref{app:convergence}. 
    
    Finally, Lemma~\ref{lem:implicit_representation} reveals a three-step construction: for a fixed parameter vector, the prediction-filtering conditions define $\hat{\mathbf{X}}(\bm{\theta})$, the smoothing conditions then define $\hat{\mathbf{S}}(\bm{\theta})$, and the parameter vector is updated by minimizing the resulting $\bm{\lambda}$-based~\gls{td} residual. This structure guides and justifies the practical implementation described in Section~\ref{sec:lambda_learning_algorithm}.

\section{The \texorpdfstring{$\bm{\lambda}$\emph{-learning}}{lambda-learning} Algorithm}
\label{sec:lambda_learning_algorithm}
        
    The local reduced formulation~\eqref{eq:reduced_min} is useful for the convergence analysis, but it cannot be used directly as a computational algorithm for two reasons. First, the reduced formulation relies on Assumptions~\ref{ass:feasibility_optimal} and~\ref{ass:smoothness_and_existence}, which, for a real-world finite-dimensional approximation, may not hold. Therefore, enforcing $\mathbf{r}_{X}(\mathbf{X};\bm{\theta})=\mathbf{0}$ and $\mathbf{r}_{S}(\mathbf{S},\mathbf{X};\bm{\theta})=\mathbf{0}$ exactly would be too restrictive for numerical training. Second, even when the local implicit representation exists, computing the exact reduced gradient would require differentiating through the coupled trajectory-reconstruction procedure, where the predicted and filtered estimates are generated forward in time, the smoothed trajectory is generated backward in time using the forward estimates, and the parameter-update residual depends on both. Hence, an exact gradient computation would require propagating sensitivities through the entire forward--backward reconstruction, which is computationally expensive and involves retaining or recomputing the intermediate quantities generated by both recursions.

    To circumvent these issues, we introduce a \emph{practical} $\bm{\lambda}$\emph{-learning} algorithm that keeps the structure suggested by the reduced formulation, but replaces the exact implicit computations with explicit trajectory-reconstruction steps. In these steps, the prediction, filtering and smoothing conditions are not imposed as exact equality constraints. Instead, consistently with the standard least-squares and quadratic-penalty treatment of nonlinear equality residuals~\cite{nocedal2006numerical,bellavia2010convergence}, they are relaxed into least-squares minimization problems to avoid feasibility issues. The reconstructed trajectories are then treated as fixed samples during the parameter-update step. Starting from an initial parameter vector $\bm{\theta}^{(0)}_{*}$, each iteration processes a sampled measurement trajectory through three operations. First, a forward estimation step computes predicted and filtered estimate trajectories. Then, a backward smoothing step computes a smoothed trajectory, using the trajectories obtained in the forward pass. Third, the parameter vector is updated by reducing the $\bm{\lambda}$-based~\gls{td} cost evaluated along these reconstructed trajectories. 

\subsubsection*{Forward Rollout}
    Since the practical algorithm processes one sampled trajectory at a time, the prediction-filtering residual $\mathbf{r}_{X}$ is replaced by its trajectory-wise counterpart $\mathbf{r}_{X}^{(i)}$, obtained by evaluating~\eqref{eq:explicit_RX} only on $\mathbf{Y}^{(i)}$. The forward rollout then approximately enforces the prediction-filtering constraint~\eqref{eq:R3eq} on the $i$-th trajectory by solving the nonlinear least-squares problem
    \begin{equation}
    \label{eq:approx_lambda_ests}
        \min_{\mathbf{X}^{(i)}} \left\|\mathbf{r}^{(i)}_{X}(\mathbf{X};\bm{\theta}^{(i-1)}_{*})\right\|^{2}_{2}.
    \end{equation}
    Here, $\mathbf{X} := [\mathbf{p}_{1}^{\top} \; \mathbf{u}_{1}^{\top} \; \cdots \, \mathbf{p}_{T}^{\top} \; \mathbf{u}_{T}^{\top}]^{\top}$ stacks the candidate predicted and filtered trajectories associated with $\mathbf{Y}^{(i)}$, while $\bm{\theta}^{(i-1)}_{*}$ denotes the parameter vector available at the beginning of the $i$-th iteration. The solution of~\eqref{eq:approx_lambda_ests}, denoted by $\hat{\mathbf{X}}^{(i)}$, provides the predicted and filtered estimate trajectories. It can be interpreted as a computable approximation of the implicit function $\hat{\mathbf{X}}(\bm{\theta}^{(i-1)}_{*})$ introduced in Lemma~\ref{lem:implicit_representation}. Although~\eqref{eq:approx_lambda_ests} is written as a trajectory-level least-squares problem, the residual $\mathbf{r}_{X}^{(i)}(\cdot)$ inherits the temporal block structure of $\mathbf{r}_{X}(\cdot)$. Therefore, the forward rollout may be implemented either over the full trajectory or sequentially up to time $T$, starting from the initial prior information and recursively computing the filtered and predicted estimates.

\subsubsection*{Backward Rollout}
    Given the result of the forward rollout, the backward rollout computes the associated smoothed trajectory by approximately enforcing the smoothing constraint~\eqref{eq:R2eq}. As in the forward rollout, the residual $\mathbf{r}_{S}$ is replaced by its trajectory-wise counterpart $\mathbf{r}_{S}^{(i)}$, obtained by evaluating~\eqref{eq:explicit_RS} only on $\mathbf{Y}^{(i)}$. The backward rollout is then formulated as
    \begin{equation}
    \label{eq:min_backward}
        \min_{\mathbf{S}^{(i)}} \left\|\mathbf{r}_{S}^{(i)}(\mathbf{S}^{(i)}, \hat{\mathbf{X}}^{(i)};\bm{\theta}^{(i-1)}_{*})\right\|^{2}_{2},
    \end{equation}
    where $\mathbf{S} = [\mathbf{s}_{1}^{\top} \; \cdots\;  \mathbf{s}_{T}^{\top}]^{\top}$ collects the candidate smoothed trajectory. The solution of~\eqref{eq:min_backward}, denoted by $\hat{\mathbf{S}}^{(i)}$, yields the smoothed trajectory and provides a computable approximation of the implicit function $\hat{\mathbf{S}}(\bm{\theta}^{(i-1)}_{*})$, evaluated along the reconstructed prediction-filtering trajectory $\hat{\mathbf{X}}^{(i)}$. This trajectory can be computed either by solving~\eqref{eq:min_backward} as a single least-squares problem or sequentially backward in time from a prescribed terminal term $\overline{\mathbf{s}}_{T+1}^{(i)}$, which initializes the backward smoothing rollout. A natural deterministic choice is to set it equal to the terminal predicted estimate obtained from the forward rollout, namely $\hat{\mathbf{x}}_{T+1|T}^{(i)}$. Here, instead, we sample according to
    \begin{equation}
    \label{eq:terminal_exploration}
        \overline{\mathbf{s}}_{T+1}^{(i)} = \hat{\mathbf{x}}_{T+1|T}^{(i)} + \bm{\xi}^{(i)}.
    \end{equation}
    This perturbation acts as an exploration mechanism for the backward rollout. By generating smoothed trajectories associated with nearby terminal states, it enriches the state locations at which the $\bm{\lambda}$-based residual is evaluated during the parameter update. In the numerical implementation, $\bm{\xi}^{(i)}$ is chosen as a zero-mean Gaussian perturbation with covariance $\bm{\Sigma}_{\mathrm{exp}}$, which controls the size and direction of the terminal exploration.

\subsubsection*{Parameter Update}
    The parameter update concludes the $i$-th iteration. Given the sampled measurement trajectory and the rollout trajectories $\hat{\mathbf{X}}^{(i)}$ and $\hat{\mathbf{S}}^{(i)}$ computed in the previous phases, the parameter vector is updated by performing only one gradient-based step on the regularized nonlinear least-squares problem
    \begin{equation}
    \label{eq:min_par_update}
        \min_{\bm{\theta}} \left\| \bm{\delta}^{(i)} \big(\hat{\mathbf{S}}^{(i)}, \hat{\mathbf{X}}^{(i)}; \bm{\theta}\big)\right\|^{2}_{2} + \rho \|\bm{\theta}\|^{2}_{2},
    \end{equation}
    where $\bm{\delta}^{(i)}$ denotes the $\bm{\lambda}$-based~\gls{td} residual evaluated on the sampled trajectory $\mathbf{Y}^{(i)}$. Problem~\eqref{eq:min_par_update} is a sample-based surrogate of the reduced objective~\eqref{eq:psi}: the implicit trajectories $\hat{\mathbf{X}}(\bm{\theta})$ and $\hat{\mathbf{S}}(\bm{\theta})$ are replaced by the rollout trajectories computed with the current parameter vector and are treated as fixed during the parameter update. Thus, the update does not use the full implicit gradient of~\eqref{eq:psi}; instead, it fits $\tilde{\bm{\lambda}}$ to the gradient targets induced by~\eqref{eq:lambda_from_ell} along the reconstructed trajectories. In the terminology of~\cite[Ch.~9]{sutton2018reinforcement}, this is a semi-gradient update of~\eqref{eq:psi}. Together with the inexact rollouts, the single-trajectory sampling, and the perturbed terminal vector~\eqref{eq:terminal_exploration}, this places Algorithm~\ref{alg:lambda_total} outside the scope of the convergence analysis in Appendix~\ref{app:convergence}. The resulting scheme can instead be interpreted as an inexact block-coordinate implementation of the ideal reduced problem: the rollouts approximate the implicit functions, while the parameter update approximates the minimization of the~\gls{td} objective.
    \begin{algorithm}[t]
    \caption{$\bm{\lambda}$\emph{-learning}}
    \label{alg:lambda_total}
        \begin{algorithmic}[1]
            \State Initialize parameters $\bm{\theta}^{(0)}_{*}$
            \State Initialize iteration index $i = 1$
            \While{stopping criterion is not met \textbf{and} $i \leq N_{\textrm{max}}$}
                \State Sample a trajectory $\mathbf{Y}^{(i)}$ from $\mathcal{D}$
                \State \textbf{Forward rollout:}
                \For{$t = 1$ to $T$}
                    \State Compute $\hat{\mathbf{x}}_{t|t}^{(i)}$ and $\hat{\mathbf{x}}_{t+1|t}^{(i)}$ from~\eqref{eq:approx_lambda_ests}
                \EndFor
                \State \textbf{Backward rollout:}
                \State Set $\overline{\mathbf{s}}_{T+1}^{(i)}$ using~\eqref{eq:terminal_exploration}
                \For{$t = T$ down to $1$}
                    \State Compute smoothed estimate $\hat{\mathbf{x}}_{t|T}^{(i)}$ from~\eqref{eq:min_backward}
                \EndFor
                \State \textbf{Parameter update:}
                \State Set $\bm{\theta}^{(i)}_{*}$ by taking a gradient-based step on~\eqref{eq:min_par_update}
                \State $i \leftarrow i + 1$
            \EndWhile
        \end{algorithmic}
    \end{algorithm}

    Once the parameter vector $\bm{\theta}_{*}$ has been obtained, the learned function $\tilde{\bm{\lambda}}$ defines both a recursive estimator and a smoother. Recursive state estimation is performed online by executing the forward rollout sequentially with fixed parameters $\bm{\theta}_{*}$. Since this operation does not require a fixed terminal horizon, it can be applied to sequences of arbitrary length. Smoothing is instead performed offline over a finite horizon by executing the backward rollout, initialized from the terminal predicted estimate of the forward rollout, $\overline{\mathbf{s}}_{T+1}=\hat{\mathbf{x}}_{T+1|T}$, without exploration noise. Thus, the same learned $\bm{\lambda}$-function is used for both recursive estimation and smoothing tasks.

\section{Numerical Experiments}
\label{sec:numerical_experiments}

    This section reports two numerical experiments assessing the proposed practical $\bm{\lambda}$\emph{-learning} algorithm. First, we run simulations with a linear system and Gaussian noise to verify that the method recovers the steady-state information matrix of the Kalman filter. Then, we simulate a nonlinear system and compare it against Bayesian and optimization-based baselines in terms of estimation accuracy, consistency, and computational cost.

\subsection{Linear Case}

    We consider the system
    \begin{align*}
        \mathbf{x}_{t+1} &= \begin{bmatrix}
            1 & 0.6 \\
            0 & 1
        \end{bmatrix} \mathbf{x}_{t} + \mathbf{w}_{t},  \quad
        \mathbf{y}_{t} = \begin{bmatrix}
            1 & 0.5
        \end{bmatrix} \mathbf{x}_{t} + \mathbf{v}_{t},
    \end{align*}
    where $\mathbf{x}_{t}, \mathbf{w}_{t} \in \mathbb{R}^{2}$ and $\mathbf{y}_{t}, \mathbf{v}_{t} \in \mathbb{R}$. The process and measurement noises are independent Gaussian white sequences with covariances $\mathbf{Q} = \operatorname{diag}(0.36, 6.25\cdot10^{-2})$ and $\mathbf{R} = 1.44$, respectively. The initial prior is $\mathbf{x}_1 \sim \mathcal{N}(\mathbf{m}_1,\mathbf{P})$ with $\mathbf{m}_1 = [0 \; 0.1]^{\top}$ and $\mathbf{P}$ chosen as the stabilizing solution of the discrete-time algebraic Riccati equation, i.e., the steady-state predicted covariance, so that the exact $\bm{\lambda}$-function is time-invariant. Here the optimal recursive estimator and smoother are the~\gls{kf} and~\gls{rtss}~\cite{rauch1965maximum}, and the optimal value function is quadratic in the prediction error $(\mathbf{s}_{t+1}-\hat{\mathbf{x}}_{t+1|t})$, if the stage and initial costs are quadratic as in~\eqref{eq:stage_cost} and~\eqref{eq:initial_time_cost}. Thus, the associated $\bm{\lambda}$-function is linear with weighting matrix $\mathbf{P}^{-1}$. To verify the recovery of $\mathbf{P}^{-1}$, we parametrize $\tilde{\bm{\lambda}}(\mathbf{s}_{t+1}, \hat{\mathbf{x}}_{t+1|t};\bm{\theta}) = \bm{\Lambda}(\bm{\theta})(\mathbf{s}_{t+1} - \hat{\mathbf{x}}_{t+1|t})$, with $\bm{\Lambda}(\bm{\theta}) = \mathbf{L}(\bm{\theta})\, \mathbf{L}^{\top}(\bm{\theta})$ (as in~\cite{masti2021tuning}), where $\mathbf{L}(\bm{\theta})$ is lower triangular. Strict positive definiteness of $\bm{\Lambda}(\bm{\theta})$ is ensured by parametrizing the diagonal entries of $\mathbf{L}(\bm{\theta})$ as $[\mathbf{L}(\bm{\theta})]_{ii} = \operatorname{softplus}(\theta_{ii})+\epsilon$, with $\epsilon>0$. The remaining lower-triangular entries are directly parametrized by the corresponding components of $\bm{\theta}$.
    
\begin{figure}[t]
        \centering
        \includegraphics[width=0.6\linewidth, trim={0.31cm 0.26cm 0.27cm 0.25cm}, clip]{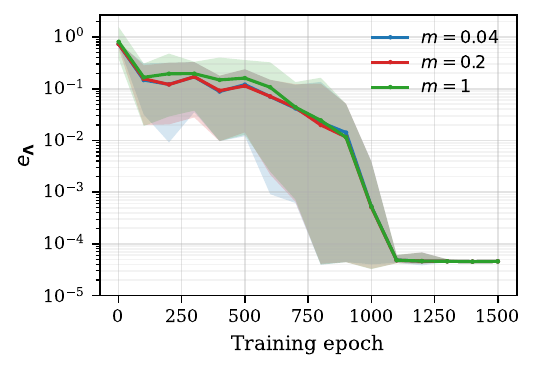}
        \caption{Linear case. Evolution of $e_{\bm{\Lambda}}$ on the validation set, reported for different values of the initialization scale $m$.}
        \label{fig:lambda_convergence}
    \end{figure}

    The dataset $\mathcal{D}$ is split into $1000$ training, $200$ validation, and $1000$ test trajectories, used respectively to update $\bm{\theta}$, select hyperparameters, and perform the final evaluation. The terminal vectors $\overline{\mathbf{s}}_{T+1}$ are generated as in~\eqref{eq:terminal_exploration}, with exploration covariance $\bm{\Sigma}_{\textrm{exp}}=10^{4}\cdot\mathbf{I}_{n_{x}}$. Robustness is assessed over ten simulations per initialization scale $m\in\{0.04,0.2,1\}$, sampling the entries of $\bm{\theta}$ from $\mathcal{N}(0,m^{2})$. The rollout least-squares subproblems are solved with amsgrad~\cite{reddi2019convergence}, while $\bm{\theta}$ is updated by~\gls{gd} with regularization weight $\rho=5\cdot10^{-5}$. Table~\ref{tab:linear_results} reports the normalized squared residual norms of the $\bm{\lambda}$-recursion at initialization and after training, on test-set trajectories generated by $\bm{\lambda}$-\emph{learning}. Since the residual blocks have different dimensions, the squared norms are normalized by the number of scalar residual equations at each time instant. Despite the large variability at initialization, especially for initialization scale $m=1$, the final residuals are close to zero and numerically indistinguishable across the considered values of $m$; hence, a single final row is reported. The optimality residuals $\mathbf{r}_S$ and $\mathbf{r}_X$ remain at solver tolerance throughout, whereas the~\gls{td} residual $\bm{\delta}$ decreases by about nine orders of magnitude. The comparatively large initial residuals for $m=1$ are driven by one markedly atypical run.
    \begin{table}[t]
        \centering
        \caption{Linear case. Mean $\pm$ standard deviation of the \emph{initial} and \emph{final} normalized squared residual norms evaluated on the test-set estimate trajectories generated by $\bm{\lambda}$-\emph{learning}.}
        \label{tab:linear_results}
        \small
        \setlength{\tabcolsep}{6pt}
        \begin{tabular}{ccccc}
            \toprule
            $m$ & Stage & $\|\bm{\delta}\|^{2}/n_{x}$ & $\|\mathbf{r}_{S}\|^{2}/n_{x}$ & $\|\mathbf{r}_{X}\|^{2}/(2n_{x})$ \\
            \midrule
            $0.04$
            & Initial
            & $0.28 \pm 0.02$
            & $(1.34 \pm 0.03)\cdot 10^{-9}$
            & $(2.04 \pm 0.15)\cdot 10^{-10}$ \\
            
            $0.2$
            & Initial
            & $0.33 \pm 0.08$
            & $(1.39 \pm 0.18)\cdot 10^{-9}$
            & $(2.55 \pm 1.53)\cdot 10^{-10}$ \\
            
            $1$
            & Initial
            & $(2.19 \pm 6.56)\cdot 10^{6}$
            & $(0.73 \pm 4.35)\cdot 10^{9}$
            & $(5.93 \pm 1.1)\cdot 10^{6}$ \\
            \midrule
            
            \textrm{--}
            & Final
            & $(3.08 \pm 0.03)\cdot 10^{-10}$
            & $(1.421 \pm 0.001)\cdot 10^{-9}$
            & $(2.07 \pm 0.01)\cdot 10^{-10}$ \\
            \bottomrule
        \end{tabular}
    \end{table}
    Moreover, Figure~\ref{fig:lambda_convergence} shows the evolution of the relative Frobenius error
    \begin{equation*}
        e_{\bm{\Lambda}} =
        \|\bm{\Lambda}(\bm{\theta}) - \mathbf{P}^{-1}\|_{F}
        / \|\mathbf{P}^{-1}\|_{F},
    \end{equation*}
    on the validation set. The learned matrix approaches the information matrix regardless of the initialization scale, additionally validating the proposed framework.

\subsection{Nonlinear Second-Order Motion in a Vertical Plane}
    We consider an autonomous nonlinear system describing the second-order motion of an object in a vertical plane with state and transition map
    \begin{align*}
        \mathbf{x}_{t} &=
        \begin{bmatrix}
            x_{t} \\ v_{x,t} \\ y_{t} \\ v_{y,t} \\ \beta_{t} \\ \omega_{t}
        \end{bmatrix},
        ~~
        \mathbf{f}(\mathbf{x}_{t}) =
        \begin{bmatrix}
            x_{t} + \Delta t\, v_{x,t} \\
            v_{x,t} + \Delta t\, a_{t}\cos(\beta_{t}) \\
            y_{t} + \Delta t\, v_{y,t} \\
            v_{y,t} + \Delta t\, a_{t}\sin(\beta_{t}) \\
            \beta_{t} + \Delta t\, \omega_{t} \\
            \omega_{t} + \Delta t\, \alpha_{t}
        \end{bmatrix}.\\ \text{with}\\
            \alpha_{t} &= -1 - 2\sin(\beta_{t}) - 0.6\,\omega_{t}, \\
            a_{t} &= \frac{-1.2\cos(\beta_{t}) -1 +0.6\sqrt{v_{x,t}^{2}+v_{y,t}^{2}+\epsilon}}{\sqrt{x_{t}^{2}+y_{t}^{2}+1}}
    \end{align*}
    Here $(x_t,y_t)$ denotes the Cartesian position, $(v_{x,t},v_{y,t})$ are the velocity components, $\beta_t$ is the orientation angle, and $\omega_t$ is the angular velocity. We set the time step to $\Delta t=0.2$ and $\epsilon = 10^{-2}$. The measured output $\mathbf{y}_{t}\in\mathbb{R}^{7}$ is generated according to
    \begin{equation*}
        \mathbf{h}(\mathbf{x}_{t})=
        \begin{bmatrix}
            \displaystyle \frac{1}{x_t^2+y_t^2+\epsilon} \\[1ex]
            \displaystyle 10\tanh\!\left(\frac{x_t v_{x,t}+y_t v_{y,t}}{10\sqrt{x_t^2+y_t^2+\epsilon}}\right)\\[1ex]
            \displaystyle \frac{x_t}{\sqrt{x_t^2+y_t^2+\epsilon}}\\[1ex]
            \displaystyle \exp\!\left(-\frac{y_t}{5}\right)\\[1ex]
            \displaystyle \frac{x_t\cos(\beta_t)+(y_t-0.5)\sin(\beta_t)}{\sqrt{x_t^2+(y_t-0.5)^2+\epsilon}}\\[1ex]
            \displaystyle \frac{-x_t\sin(\beta_t)+(y_t-0.5)\cos(\beta_t)}{\sqrt{x_t^2+(y_t-0.5)^2+\epsilon}}\\[1ex]
            \displaystyle \omega_t+0.08\omega_t^3
        \end{bmatrix}.
    \end{equation*}
    The process and measurement noises, $\mathbf{w}_t\in\mathbb{R}^{6}$ and $\mathbf{v}_t\in\mathbb{R}^{7}$, are mutually independent, zero-mean multivariate Student-$t$ random variables with $\nu=3$ degrees of freedom and covariance matrices 
    \begin{align*}
        \mathbf{Q} &= \operatorname{diag}(0.02,\,0.08,\,0.02,\,0.08,\,0.04,\,0.1),\\
        \mathbf{R} &= \operatorname{diag}(0.08,\,0.05,\,0.015,\,0.015,\,0.07,\,0.07,\,0.02).
    \end{align*}
    The initial state is distributed according to $\mathbf{x}_{1} \sim \mathcal{N}(\mathbf{m}_{1},\mathbf{P}_{1})$, with $ \mathbf{m}_{1}=[0.25,-1,0.5,0.4,1.2,-1]^\top$ and $\mathbf{P}_{1} = \operatorname{diag}(0.6,\,0.4,\,0.6,\,0.4,\,0.4,\,0.3)$. 
    
    For the $\bm{\lambda}$-function, we choose a parametrization that is linear in the prediction error, but with an associated information matrix that depends nonlinearly on the predicted estimate. Specifically, we consider $\tilde{\bm{\lambda}}(\mathbf{s}_{t+1}, \hat{\mathbf{x}}_{t+1|t};\bm{\theta}) = \bm{\Lambda}(\hat{\mathbf{x}}_{t+1|t};\bm{\theta})(\mathbf{s}_{t+1} - \hat{\mathbf{x}}_{t+1|t})$, where $\bm{\Lambda}(\hat{\mathbf{x}}_{t+1|t};\bm{\theta}) = \mathbf{L}(\hat{\mathbf{x}}_{t+1|t};\bm{\theta})\, \mathbf{L}^{\top}(\hat{\mathbf{x}}_{t+1|t};\bm{\theta})$. The matrix $\mathbf{L}$ is lower triangular, and its entries are generated by a feedforward neural network with two hidden layers of $64$ neurons and~\gls{gelu} activations. A softplus activation, shifted by $10^{-6}$, is applied to ensure positive diagonal entries and hence positive definiteness of $\bm{\Lambda}$. This parametrization can be interpreted as the gradient of a local quadratic approximation of the optimal value function, in analogy with quadratic arrival costs commonly used in~\gls{mhe}~\cite{rao2003constrained,rawlings2020stateestimation,alessandri2008moving}. We use $1000$ training, $200$ validation, and $1000$ test trajectories, each containing $40$ measurements. The terminal decision variable $\overline{\mathbf{s}}_{T+1}$ is generated according to~\eqref{eq:terminal_exploration}, with $\bm{\Sigma}_{\textrm{exp}}$ decreasing linearly from $\mathbf{I}_{n_x}$ to $10^{-2}\mathbf{I}_{n_x}$. The neural network biases are initialized to zero, whereas the weights of layer $\ell$ are sampled independently according to $W_{\ell,ij}\sim\mathcal{N}(0,m^2/n_{\ell-1})$, where $n_{\ell-1}$ is the input dimension of the layer and $m\in\{0.04,0.2,1.0\}$ is the initialization scale. For each $m$, we perform ten training runs using different training seeds on $\mathcal{D}$. Parameters are updated by~\gls{gd} using one randomly sampled trajectory per batch for $150000$ parameter updates, corresponding to $150$ nominal passes over the training set. The learning rate follows a cosine schedule from $5\cdot10^{-3}$ to $10^{-6}$. The forward and backward rollout problems in Algorithm~\ref{alg:lambda_total} are solved using \texttt{amsgrad}~\cite{reddi2019convergence}. The regularization weight is $\rho=5\cdot10^{-5}$, as in the linear case.
    
    Estimation accuracy is assessed through the~\gls{mae}, defined for trajectory $i$ and time $t$ as
    \begin{equation*}
        \mathrm{MAE}_{t}^{(i)} = \frac{1}{n_x}\left\|\hat{\mathbf{x}}_{t}^{(i)}-\mathbf{x}_{t}^{(i)}\right\|_1.
    \end{equation*}
    This metric is computed separately for predicted, filtered, and smoothed estimates over all test trajectories, time instants, and independent runs. Instead, empirical consistency is evaluated through the predicted~\gls{nees},
    \begin{equation*}
        \mathrm{NEES}_{t+1|t}^{(i)} = \frac{\left\|\hat{\mathbf{x}}_{t+1|t}^{(i)}-\mathbf{x}_{t+1}^{(i)}\right\|^{2}_{\bm{\Lambda}_{t+1|t}^{(i)}}}{n_x},
    \end{equation*}
    where $\bm{\Lambda}_{t+1|t}^{(i)}$ is interpreted as the predicted information matrix. Under this interpretation, values close to one indicate empirical consistency of the predictor, while larger values indicate overconfidence. In the present work, this metric is particularly meaningful because both the proposed method and the optimization-based baselines approximate the predicted information matrix. The computational burden is measured as average runtime per discrete time step. For filtering comparisons, only the forward rollout of the proposed method is timed, including both prediction and measurement correction. For smoothing comparisons, both the forward and backward rollouts are included. Compilation, metric evaluation, and file serialization are excluded. The proposed method and the Bayesian baselines are implemented in Python using JAX, the~\gls{mhe} baselines use the code generated by \texttt{acados}, and all runtimes are measured single-threaded on a 13th Gen Intel Core i7-13650HX CPU (2.60\,GHz).
    
    The proposed method is evaluated against several baselines: the~\gls{ekf},~\gls{ukf}, and~\gls{pf}, together with their smoothing extensions, namely the~\gls{ertss}, the~\gls{urtss}, and the backward-simulation particle smoother, all implemented according to the respective formulations described in~\cite{sarkka2023bayesian}. The particle filter is evaluated using $1500$ and $6000$ particles, while the backward-simulation particle smoother uses $M = 32$ backward samples. The optimization-based baselines are \gls{mhe} estimators implemented in \texttt{acados}~\cite{Verschueren2021}. At the beginning of each horizon, their arrival cost is $\bm{\Gamma}_{t}(\mathbf{s}_{t},\hat{\mathbf{x}}_{t|t-1})=(\mathbf{s}_{t}-\hat{\mathbf{x}}_{t|t-1})^{\top}{\bm{\Lambda}_{t|t-1}}(\mathbf{s}_{t}-\hat{\mathbf{x}}_{t|t-1})$. In the fixed (F) configuration, the arrival information matrix is set to the initial prior information matrix $\mathbf{P}_1^{-1}$ at every step. In the recursive (R) configuration, it is updated with the predicted information matrix computed at the preceding step from either the exact Hessian (H) of the~\gls{mhe} problem or its Gauss--Newton approximation~\cite{baumgartner2022moving} (GN). The~\gls{mhe} problem is solved by sequential quadratic programming (SQP). For each configuration, we report the best-performing horizon among $N_{\mathrm{MHE}}\in\{2,5,10\}$. The method labels in Table~\ref{tab:vertical_plane_estimators} identify these choices; for example, MHE H-SQP-R2 uses the exact Hessian, SQP, the recursive arrival matrix, and a horizon of two.
    \begin{table}[t]
    \caption{Vertical-plane case. Recursive estimator performance. The Pred. and Filt. columns report $10\cdot\mathrm{MAE}$, while \gls{nees} is reported without scaling. Metrics are reported as mean $\pm$ standard deviation over all test trajectories and time instants, pooling the ten training runs for $\lambda$-L. Execution time is the average per-step runtime.}
    \label{tab:vertical_plane_estimators}
    \centering
    \small
    \setlength{\tabcolsep}{6pt}
    \begin{tabular}{lcccc}
    \toprule
    Method
    & Pred. $\mathrm{MAE}\times 10$
    & Filt. $\mathrm{MAE}\times 10$
    & NEES
    & Time [ms] \\
    \midrule
    EKF
    & $5.83{\pm}11.21$
    & $5.09{\pm}10.90$
    & $20.80{\pm}136.77$
    & $0.012$ \\
    UKF
    & $5.31{\pm}10.56$
    & $4.57{\pm}10.30$
    & $9.50{\pm}91.34$
    & $0.009$ \\
    PF$_{1500}$
    & $4.56{\pm}7.94$
    & $3.78{\pm}7.66$
    & $52.24{\pm}1783.99$
    & $0.564$ \\
    PF$_{6000}$
    & $4.11{\pm}8.11$
    & $3.33{\pm}7.79$
    & $38.56{\pm}1264.60$
    & $1.249$ \\
    MHE H-SQP-R2
    & $4.12{\pm}3.95$
    & $3.34{\pm}3.73$
    & $6.10{\pm}37.84$
    & $1.455$ \\
    MHE H-SQP-F10
    & $4.18{\pm}4.47$
    & $3.40{\pm}4.26$
    & $6.65{\pm}44.67$
    & $8.123$ \\
    MHE GN-SQP-R2
    & $4.76{\pm}4.64$
    & $3.99{\pm}4.47$
    & $11.04{\pm}38.74$
    & $0.981$ \\
    MHE GN-SQP-F10
    & $4.79{\pm}4.63$
    & $4.03{\pm}4.46$
    & $11.26{\pm}38.90$
    & $2.879$ \\
    \midrule
    $\lambda$-L$_{1}$
    & $4.33{\pm}4.30$
    & $3.55{\pm}4.09$
    & $6.52{\pm}39.64$
    & $0.342$ \\
    $\lambda$-L$_{0.2}$
    & $4.35{\pm}4.65$
    & $3.57{\pm}4.42$
    & $7.26{\pm}116.37$
    & $0.364$ \\
    $\lambda$-L$_{0.04}$
    & $4.36{\pm}4.39$
    & $3.58{\pm}4.18$
    & $6.71{\pm}40.78$
    & $0.338$ \\
    \bottomrule
    \end{tabular}
    \end{table}
    \begin{table}[t]
    \caption{Vertical-plane case. Smoother performance. The MAE column reports $10\cdot\mathrm{MAE}$, while runtime is the average per-step execution time in milliseconds. Metrics are reported as mean $\pm$ standard deviation over all test trajectories and time instants, pooling the ten training runs for $\lambda$-L.}
    \label{tab:vertical_plane_smoothers}
    \centering
    \small
    \setlength{\tabcolsep}{5pt}
    \renewcommand{\arraystretch}{1.08}
    \begin{tabular}{lcc}
    \toprule
    Method & $\mathrm{MAE}\times 10$ & Time [ms] \\
    \midrule
    ERTSS              & $4.38{\pm}10.60$ & $0.016$ \\
    URTSS              & $3.77{\pm}9.75$  & $0.014$ \\
    BSPS$_{1500}$      & $3.35{\pm}7.61$  & $1.253$ \\
    BSPS$_{6000}$      & $2.85{\pm}7.74$  & $3.467$ \\
    \midrule
    $\lambda$-L$_{1}$    & $2.97{\pm}3.86$ & $0.567$ \\
    $\lambda$-L$_{0.2}$  & $2.98{\pm}4.16$ & $0.576$ \\
    $\lambda$-L$_{0.04}$ & $2.99{\pm}3.95$ & $0.577$ \\
    \bottomrule
    \end{tabular}
    \end{table}
    Tables~\ref{tab:vertical_plane_estimators} and~\ref{tab:vertical_plane_smoothers} report estimation and smoothing results. The notation $\lambda$-L$_m$ in the tables denotes the proposed $\bm{\lambda}$-\emph{learning} method with initialization scale $m$. 

    Overall, the results position the $\bm{\lambda}$-parametrization as a favorable trade-off between accuracy, consistency, and computational cost. The~\gls{ekf} and~\gls{ukf} are the cheapest estimators but yield the least accurate estimates and large predicted~\gls{nees}, while the~\gls{pf} and the~\gls{mhe} baselines improve accuracy only at a higher computational cost, through more particles, longer horizons, or exact-Hessian formulations. The proposed method attains accuracy comparable to the best-performing baselines, MHE H-SQP-R2 and PF$_{6000}$, and predicted~\gls{nees} comparable to MHE H-SQP-R2, while running about four times faster. It is also more accurate and better calibrated than Gauss--Newton MHE at about one third of its runtime. This suggests that the learned information matrix amortizes offline part of the computational effort that classical methods spend online at every step, through resampling, particle propagation, or the solution of an optimization problem. For smoothing, the method is more accurate than BSPS$_{1500}$ at less than half of its runtime, approaches the accuracy of BSPS$_{6000}$ at about one sixth of its runtime, and clearly outperforms the Rauch--Tung--Striebel-type smoothers, which are cheaper but less accurate. These results are robust to the initialization scale $m$, which supports the use of the method in applications where both accuracy and computational cost matter, such as real-time estimation. The large standard deviations, common to all methods, reflect the heavy-tailed Student-$t$ noise, whose rare large disturbances produce occasional large estimation errors. Both the~\gls{mhe} baselines and the proposed method use quadratic stage costs, which are not matched to this noise distribution. The results show that this choice still yields accurate estimates, while making all methods sensitive to these rare events, as reflected by standard deviations that exceed the corresponding means.
    
\section{Conclusions}
\label{sec:conclusions}
    In this work, we proposed a learning-based framework for recursive state estimation and smoothing grounded in the~\gls{dp} structure of the~\gls{fie} problem. We introduced the $\bm{\lambda}$-function, defined as the gradient of the optimal value function, and showed that it satisfies a forward recursion in which past information enters only implicitly, through optimality conditions. This recursion yields a~\gls{td} learning problem whose targets involve no evaluation of the learned function and require no ground-truth states, and from which both a recursive estimator and a smoother are obtained. For an idealized version of this problem, we established local convergence of gradient descent to a critical point, and we proposed a practical semi-gradient algorithm, termed $\bm{\lambda}$\emph{-learning}. In the linear--Gaussian case, the method recovers the steady-state Riccati solution. In a nonlinear benchmark with heavy-tailed noise, it attains accuracy comparable to the best Bayesian and optimization-based baselines and predicted~\gls{nees} comparable to the best~\gls{mhe} baseline, while running about four times faster, and it approaches the accuracy of the best particle smoother at about one sixth of its runtime. Future work will extend the convergence analysis to account for approximate inner optimizations and the practical rollout-based implementation. It will also investigate richer parametrizations of the $\bm{\lambda}$-function for complex or high-dimensional nonlinear systems.

\bibliographystyle{plain}
\bibliography{references_autart}

\appendix
\section{Convergence Analysis of the Reduced Learning Problem}
\label{app:convergence}

    This appendix studies the local convergence properties of~\gls{gd} applied to the reduced unconstrained problem~\eqref{eq:reduced_min}. The reduced problem is smooth and generally nonconvex. The convergence analysis proceeds in two steps. First, sufficient decrease and bounded-gradient properties are established for~\gls{gd}, yielding convergence to the set of critical points. Then, by invoking the~\gls{kl} property, the result is strengthened to convergence of the whole sequence to a single critical point~\cite{attouch2013convergence,bolte2014proximal}.

    We first establish the smoothness of the cost $\Psi$ in~\eqref{eq:psi}.
    \begin{lemma}
    \label{lemma:psi_differentiability}
        The reduced cost $\Psi(\bm{\theta})$ is continuously differentiable on $\mathcal{U}_{\theta}$.
    \end{lemma}
    \begin{proof}
        The result follows from the chain rule, Assumption~\ref{ass:smoothness_and_existence}, and Lemma~\ref{lem:implicit_representation}, since $\bm{\delta}$ and the implicit functions $\hat{\mathbf{S}}$ and $\hat{\mathbf{X}}$ are continuously differentiable on $\mathcal{U}_{\theta}$.
    \end{proof}
    However, to ensure convergence of~\gls{gd} updates applied to $\Psi(\bm{\theta})$, we require the following standard regularity assumption.
    \begin{assumption}[Lipschitz continuity of the gradient]
    \label{ass:lipschitz_continuity}
        The gradient $\nabla_{\bm{\theta}} \Psi(\bm{\theta})$ is Lipschitz continuous on $\mathcal{U}_{\theta}$ with modulus $L_{\bm{\theta}}$, that is, for all $\bm{\theta}_1, \bm{\theta}_2 \in \mathcal{U}_{\theta}$
        \begin{equation*}
            \|\nabla_{\bm{\theta}}\Psi(\bm{\theta}_{1})-\nabla_{\bm{\theta}}\Psi(\bm{\theta}_{2})\| \leq L_{\bm{\theta}}\,\|\bm{\theta}_1-\bm{\theta}_2\|.    
        \end{equation*}
    \end{assumption}
    Assumption~\ref{ass:lipschitz_continuity} imposes a stronger regularity property than the continuous differentiability established in Lemma~\ref{lemma:psi_differentiability}. In particular, it requires the gradient of the reduced cost to vary at most linearly with the parameter vector on $\mathcal{U}_{\theta}$. In practical terms, this excludes parametrizations for which small variations of $\bm{\theta}$ may produce arbitrarily large variations in the gradient of the reduced cost, and ensures that the local curvature of $\Psi$ remains controlled on $\mathcal{U}_{\theta}$. 
    
    Moreover, since $\Psi$ is defined through the implicit functions $\hat{\mathbf{S}}(\bm{\theta})$ and $\hat{\mathbf{X}}(\bm{\theta})$, which exist only locally, the convergence analysis applies only as long as the iterates remain in $\mathcal{U}_{\theta}$. We formalize this requirement through the following definition and assumption.
    \begin{definition}
        For a nonempty set $A\subseteq\mathbb{R}^{n_{\theta}}$ and a set $B\subseteq\mathbb{R}^{n_{\theta}}$, we define the distance between $A$ and $B$ as
        \begin{equation*}
            \operatorname{dist}(A,B) :=
            \begin{cases}
                \displaystyle \inf_{\bm{a}\in A,\bm{b}\in B}\|\bm{a}-\bm{b}\|, & B\neq\emptyset, \\
                +\infty, & B=\emptyset.
            \end{cases}
        \end{equation*}
        If $A=\{\bm{\theta}\}$ is a singleton, we use the shorthand $\operatorname{dist}(\bm{\theta},B):= \operatorname{dist}(\{\bm{\theta}\},B)$.
    \end{definition}

    \begin{assumption}
    \label{ass:level_set}
        Let $\bm{\theta}_{0}\in\mathcal{U}_{\theta}$ be the initial parameter vector and define the initial level set $\mathcal{L}_{0} := \{\bm{\theta}\in\mathcal{U}_{\theta}:\Psi(\bm{\theta})\leq\Psi(\bm{\theta}_{0})\}$. There exists $d_{0}>0$ such that $\operatorname{dist}\left(\mathcal{L}_{0},\mathbb{R}^{n_{\theta}}\setminus\mathcal{U}_{\theta}\right) \geq d_{0}$.
    \end{assumption}

    Assumption~\ref{ass:level_set} plays the role of the boundedness conditions commonly imposed in convergence analyses of descent methods~\cite{bolte2014proximal}. In addition, the margin $d_0$, which is specific to the present local reformulation, ensures that sufficiently small gradient steps from $\mathcal{L}_{0}$ remain in $\mathcal{U}_{\theta}$, where $\Psi$ is well defined. It holds trivially when the implicit functions of Lemma~\ref{lem:implicit_representation} are globally defined, i.e., $\mathcal{U}_{\theta}=\mathbb{R}^{n_\theta}$.
    \begin{lemma}
    \label{lem:compactness_and_G0}
        The initial level set $\mathcal{L}_{0}$ is compact. Moreover, there exists $G_{0}>0$ such that $\|\nabla_{\bm{\theta}}\Psi(\bm{\theta})\| \leq G_{0}$ for all $\bm{\theta}\in\mathcal{L}_{0}$.
    \end{lemma}
    \begin{proof}
        Since $\rho>0$, the reduced objective $\Psi$ in~\eqref{eq:psi} satisfies $\Psi(\bm{\theta}) \geq \rho \|\bm{\theta}\|^{2}$ for all $\bm{\theta} \in \mathcal{U}_{\theta}$. Then, for all $\bm{\theta}\in\mathcal{L}_{0}$, $\rho \|\bm{\theta}\|^{2} \leq \Psi(\bm{\theta})\leq \Psi(\bm{\theta}_{0})$, so $\|\bm{\theta}\| \leq \sqrt{\Psi(\bm{\theta}_{0})/\rho}$. Thus, $\mathcal{L}_{0}$ is bounded. For closedness, let $\{\bm{\theta}_{j}\}\subset\mathcal{L}_{0}$ with $\bm{\theta}_{j}\to\bar{\bm{\theta}}$. By Assumption~\ref{ass:level_set}, $\operatorname{dist}(\bm{\theta}_{j},\mathbb{R}^{n_{\theta}}\setminus\mathcal{U}_{\theta})\geq d_{0}$ for all $j$. Since $\bm{\theta}_{j}\to\bar{\bm{\theta}}$, this forces $\bar{\bm{\theta}}\in\mathcal{U}_{\theta}$ (otherwise $\operatorname{dist}(\bm{\theta}_{j},\mathbb{R}^{n_{\theta}}\setminus\mathcal{U}_{\theta})\leq\|\bm{\theta}_{j}-\bar{\bm{\theta}}\|\to 0$, a contradiction). By continuity of $\Psi$ on $\mathcal{U}_{\theta}$, $\Psi(\bar{\bm{\theta}})=\lim_{j}\Psi(\bm{\theta}_{j})\leq\Psi(\bm{\theta}_{0})$, so $\bar{\bm{\theta}}\in\mathcal{L}_{0}$. Thus $\mathcal{L}_{0}$ is closed, hence compact. 
        
        Finally, $\|\nabla_{\bm{\theta}}\Psi(\bm{\theta})\|$ is continuous on the compact set $\mathcal{L}_{0}$ by Lemma~\ref{lemma:psi_differentiability}. Hence, by the Weierstrass extreme value theorem~\cite{bertsekas1999nonlinear}, it attains a maximum on $\mathcal{L}_{0}$, and there exists $G_{0}>0$ such that $\|\nabla_{\bm{\theta}}\Psi(\bm{\theta})\|\le G_{0}$ for all $\bm{\theta}\in\mathcal{L}_{0}$.
    \end{proof}
    We can now establish the descent properties of~\gls{gd}.
    \begin{lemma}
    \label{lemma:sufficient_decrease}
        Let ${\bm{\theta}}_{0} \in \mathcal{U}_{\theta}$ and let $\{\bm{\theta}_{k}\}_{k \in \mathbb{N}}$ be generated by the~\gls{gd} step $\bm{\theta}_{k} = \bm{\theta}_{k-1} - \alpha_{k}\nabla_{\bm{\theta}}\Psi(\bm{\theta}_{k-1})$, where the learning rate $\alpha_k$ satisfies $0<\underline{\alpha}\leq\alpha_{k} \leq \overline{\alpha}$, where 
        \begin{equation}
        \label{eq:max_step_size}
            \overline{\alpha} := \min\left\{\frac{1}{L_{\bm{\theta}}},\frac{d_{0}}{2G_{0}}\right\}.
        \end{equation}
        Then, the following properties hold.
        \begin{enumerate}
            \item \label{item:suff_dec} (Sufficient Decrease) The sequence $\{\Psi(\bm{\theta}_{k})\}_{k\in\mathbb{N}}$ satisfies
            \begin{equation}
            \label{eq:sufficient_decrease_thm}
                \Psi(\bm{\theta}_{k}) + M\|\bm{\theta}_{k}-\bm{\theta}_{k-1}\|^{2} \le \Psi(\bm{\theta}_{k-1}),
            \end{equation}
            with $M:=\left(1/\overline{\alpha} \,-  L_{\bm{\theta}}/2 \right)>0$. Consequently, $\{\Psi(\bm{\theta}_{k})\}_{k\in\mathbb{N}}$ is nonincreasing.

            \item (Boundedness of the Iterates)
            The sequence $\{\bm{\theta}_{k}\}_{k\in\mathbb{N}}$ is contained in the initial level set $\mathcal{L}_{0}$ and therefore is bounded.

            \item (Vanishing Increments) 
            The sequence satisfies $\sum_{k=1}^{\infty}\|\bm{\theta}_{k}-\bm{\theta}_{k-1}\|^{2}<+\infty$, and hence $\|\bm{\theta}_{k}-\bm{\theta}_{k-1}\| \to 0$.
        \end{enumerate}
    \end{lemma}
    \begin{proof}
        The proof first shows by induction the first two results. By definition, $\bm{\theta}_{0}\in\mathcal{L}_{0}$. Assume now that $\bm{\theta}_{k-1}\in\mathcal{L}_{0}$ for some $k\in\mathbb{N}_{>0}$. By Lemma~\ref{lem:compactness_and_G0}, $\|\nabla_{\bm{\theta}}\Psi(\bm{\theta}_{k-1})\| \leq G_{0}$. Moreover, by~\eqref{eq:max_step_size}, $\alpha_{k}\leq d_{0}/(2G_{0})$, and, hence, $\|\bm{\theta}_{k}-\bm{\theta}_{k-1}\| = \alpha_{k} \|\nabla_{\bm{\theta}}\Psi(\bm{\theta}_{k-1})\| \leq {d_{0}}/{2}$.
        Since $\bm{\theta}_{k-1}\in\mathcal{L}_{0}$ and $\operatorname{dist} \left( \mathcal{L}_{0}, \mathbb{R}^{n_{\theta}}\setminus\mathcal{U}_{\theta} \right) \geq d_{0}$ by Assumption~\ref{ass:level_set}, the whole segment joining $\bm{\theta}_{k-1}$ and $\bm{\theta}_{k}$ is contained in $\mathcal{U}_{\theta}$. Indeed, every point on this segment has distance at most $d_{0}/2$ from $\bm{\theta}_{k-1}$, while every point in $\mathbb{R}^{n_{\theta}}\setminus\mathcal{U}_{\theta}$ has distance at least $d_{0}$ from $\bm{\theta}_{k-1}$. Therefore, the descent lemma (see, e.g.,~\cite{bolte2014proximal}) can be applied along this segment, since $\nabla_{\bm{\theta}}\Psi(\cdot)$ is Lipschitz continuous on $\mathcal{U}_{\theta}$ as required by Assumption~\ref{ass:lipschitz_continuity}, to obtain
        \begin{align*}
            \Psi(\bm{\theta}_{k}) &\leq \Psi(\bm{\theta}_{k-1}) \\
            &\phantom{=}+ (\bm{\theta}_{k}-\bm{\theta}_{k-1})^{\top} \nabla_{\bm{\theta}}\Psi(\bm{\theta}_{k-1}) \hspace{-1pt}+\hspace{-1pt} \frac{L_{\bm{\theta}}}{2} \|\bm{\theta}_{k}-\bm{\theta}_{k-1}\|^{2}.
        \end{align*}
        Substituting the~\gls{gd} update gives
        \begin{equation}
        \label{eq:order_with_alphak}
            \Psi(\bm{\theta}_{k}) + \left(\frac{1}{\alpha_k} - \frac{L_{\bm{\theta}}}{2}\right)\|\bm{\theta}_{k} - \bm{\theta}_{k-1}\|^2 \leq \Psi(\bm{\theta}_{k-1}) . 
        \end{equation}
        Since $0<\underline{\alpha}\leq\alpha_{k}\leq\overline{\alpha}$, it follows that $\frac{1}{\alpha_{k}}\geq \frac{1}{\overline{\alpha}}$ and, by~\eqref{eq:max_step_size}, $\frac{1}{\overline{\alpha}} \geq L_{\bm{\theta}}$. Then, 
        \begin{equation*}
            M := \frac{1}{\overline{\alpha}} - \frac{L_{\bm{\theta}}}{2} \geq L_{\bm{\theta}} - \frac{L_{\bm{\theta}}}{2} = \frac{L_{\bm{\theta}}}{2} > 0.
        \end{equation*}
        Consequently, from~\eqref{eq:order_with_alphak}, $\Psi(\bm{\theta}_{k}) + M\|\bm{\theta}_{k}-\bm{\theta}_{k-1}\|^{2} \leq \Psi(\bm{\theta}_{k-1})$, with $M>0$, which implies $\Psi(\bm{\theta}_{k}) \leq \Psi(\bm{\theta}_{k-1}) \leq \Psi(\bm{\theta}_{0})$. Thus, $\bm{\theta}_{k}\in\mathcal{L}_{0}$. This closes the induction and proves that $ \{\bm{\theta}_{k}\}_{k\in\mathbb{N}} \subseteq \mathcal{L}_{0} \subset \mathcal{U}_{\theta}$. Since $\mathcal{L}_{0}$ is compact by Lemma~\ref{lem:compactness_and_G0}, it follows that $\{\bm{\theta}_{k}\}_{k\in\mathbb{N}}$ is bounded. Finally, since $\{\Psi(\bm{\theta}_{k})\}_{k \in \mathbb{N}}$ is nonincreasing and bounded below by zero, it converges to some finite value $\Psi_{\infty}\ge0$ by the monotone convergence theorem. Summing the sufficient decrease inequality over $k$ yields
        \begin{align*}
            \sum_{k=1}^{\infty} \|\bm{\theta}_{k}-\bm{\theta}_{k-1}\|^2 &\le \frac{1}{M}\sum_{k=1}^{\infty} \left(\Psi(\bm{\theta}_{k-1}) - \Psi(\bm{\theta}_{k})\right) \\
            &= \frac{1}{M} \left(\Psi(\bm{\theta}_0) - \Psi_{\infty}\right) < \infty .
        \end{align*}
        Therefore, $\|\bm{\theta}_{k}-\bm{\theta}_{k-1}\| \to 0$.
    \end{proof}
    Lemma~\ref{lemma:sufficient_decrease} establishes the fundamental decrease property of the~\gls{gd} scheme applied to $\Psi$. It guarantees that the sequence $\{\bm{\theta}_{k}\}_{k \in \mathbb{N}}$ is bounded and remains in $\mathcal{L}_{0}$ and the associated sequence $\{\Psi(\bm{\theta}_{k})\}_{k \in \mathbb{N}}$ is nonincreasing. Furthermore, this result allows us to characterize the behavior of the gradient $\nabla_{\bm{\theta}} \Psi(\bm{\theta})$ on the sequence of iterates $\{\bm{\theta}_{k}\}_{k \in \mathbb{N}}$.
    \begin{lemma}
    \label{lemma:bounded_gradient}
        Let $\{\bm{\theta}_{k}\}_{k\in\mathbb{N}}$ be the sequence generated by the~\gls{gd} iteration. Then,
        \begin{enumerate}
            \item For all $k \in \{1, 2, \ldots \}$, it holds that 
            \begin{equation*}
                \|\nabla_{\bm{\theta}}\Psi(\bm{\theta}_{k-1})\|\le\frac{1}{\underline{\alpha}}\|\bm{\theta}_{k}-\bm{\theta}_{k-1}\|,
            \end{equation*}
            and therefore $\|\nabla_{\bm{\theta}}\Psi(\bm{\theta}_{k})\| \to 0$.
            \item Every accumulation point $\bm{\theta}_{*}$ of the sequence $\{\bm{\theta}_{k}\}_{k\in\mathbb{N}}$ is a critical point of $\Psi$, i.e., $\nabla_{\bm{\theta}}\Psi(\bm{\theta}_{*}) = \mathbf{0}$.
        \end{enumerate}
    \end{lemma}
    \begin{proof}
        By the~\gls{gd} update,
        \begin{equation*}
            \|\nabla_{\bm{\theta}}\Psi(\bm{\theta}_{k-1})\| = \frac{1}{\alpha_k}\|\bm{\theta}_{k}-\bm{\theta}_{k-1}\| \le \frac{1}{\underline{\alpha}} \|\bm{\theta}_{k}-\bm{\theta}_{k-1}\|,
        \end{equation*}
        which vanishes by Lemma~\ref{lemma:sufficient_decrease}. Since the sequence $\{\bm{\theta}_k\}_{k \in \mathbb{N}}$ is bounded, it admits accumulation points by the Bolzano--Weierstrass Theorem. Let now $\bm{\theta}_{*}$ be an arbitrary accumulation point. By definition, there exists a subsequence $\{\bm{\theta}_{k_j}\}_{j\in\mathbb{N}}$ such that $\bm{\theta}_{k_j}\to\bm{\theta}_{*}$ as $j\to\infty$. By continuity of the gradient, $\nabla_{\bm{\theta}}\Psi(\bm{\theta}_{*}) = \lim_{j\to\infty}\nabla_{\bm{\theta}}\Psi(\bm{\theta}_{k_j}) = \mathbf{0}$. 
    \end{proof}
    Lemma~\ref{lemma:sufficient_decrease} and Lemma~\ref{lemma:bounded_gradient} yield together the standard convergence properties of~\gls{gd}, that is, bounded iterates, vanishing steps, and subsequential convergence to critical points. These properties allow us to characterize the asymptotic behavior of the sequence of iterates. In particular, boundedness ensures the existence of accumulation points, while the sufficient decrease property guarantees convergence of the objective values. The following theorem formalizes these observations by describing the limit point set of the sequence $\{\bm{\theta}_k\}_{k \in \mathbb{N}}$.
    \begin{theorem}[Convergence to the limit point set]
    \label{theo:limit_set_convergence}
        The set of the accumulation points of $\{\bm{\theta}_{k}\}_{k \in \mathbb{N}}$, denoted by $\Omega(\{\bm{\theta}_{k}\})$, is contained in $\mathcal{L}_{0}$, nonempty, compact, and satisfies
        \begin{equation*}
            \lim_{k \to \infty}\operatorname{dist}(\bm{\theta}_{k}, \Omega(\{\bm{\theta}_{k}\})) =  0.
        \end{equation*}
        Moreover, $\Psi$ is constant on $\Omega(\{\bm{\theta}_{k}\})$, i.e., $\Psi(\bm{\theta}) = \Psi_{\infty}$ for all $\bm{\theta} \in \Omega(\{\bm{\theta}_{k}\})$.
    \end{theorem}
    \begin{proof}
        By the Bolzano--Weierstrass Theorem $\Omega(\{\bm{\theta}_{k}\})$ is nonempty. Moreover, because $\{\bm{\theta}_k\}_{k\in\mathbb{N}}\subseteq \mathcal{L}_0$ and $\mathcal{L}_0$ is compact, every accumulation point belongs to $\mathcal{L}_0$, therefore $\Omega(\{\bm{\theta}_k\}) \subseteq \mathcal{L}_0$ and $\Omega(\{\bm{\theta}_k\})$ is bounded. Since the limit point set of a bounded sequence is closed, $\Omega(\{\bm{\theta}_k\})$ is also compact. We now prove $\operatorname{dist}(\bm{\theta}_k,\Omega(\{\bm{\theta}_k\})) \to 0$ by contradiction. Suppose that there exist ${\varepsilon>0}$ and a subsequence $\{\bm{\theta}_{k_j}\}_{j \in \mathbb{N}}$ such that $\operatorname{dist}(\bm{\theta}_{k_j},\Omega(\{\bm{\theta}_k\})) \ge \varepsilon$ for all $j\in\mathbb{N}$. Since the sequence $\{\bm{\theta}_{k_j}\}_{j \in \mathbb{N}}$ is bounded, it admits a convergent subsequence $\{\bm{\theta}_{k_{j_{\ell}}}\}_{\ell \in \mathbb{N}}$ with limit $\bar{\bm{\theta}}$. By definition, $\bar{\bm{\theta}}\in \Omega(\{\bm{\theta}_k\})$. Hence $\operatorname{dist}(\bm{\theta}_{k_{j_\ell}},\Omega(\{\bm{\theta}_k\})) \leq \|\bm{\theta}_{k_{j_\ell}}-\bar{\bm{\theta}}\| \to 0$, which contradicts the inequality above. Finally, we show that $\Psi$ is constant on $\Omega(\{\bm{\theta}_k\})$. Since $\{\Psi(\bm{\theta}_k)\}$ is nonincreasing and bounded below, it converges to some value $\Psi_{\infty}$. Let $\bm{\theta}$ be an arbitrary element of $\Omega(\{\bm{\theta}_k\})$. There exists a subsequence $\{\bm{\theta}_{k_j}\}$ such that $\bm{\theta}_{k_j} \to \bm{\theta}$. By continuity of $\Psi$, $\Psi(\bm{\theta}) = \lim_{j\to\infty}\Psi(\bm{\theta}_{k_j}) = \Psi_{\infty}$. Since $\bm{\theta}$ is arbitrary, the result holds for all elements of $\Omega(\{\bm{\theta}_k\})$.
    \end{proof}
    Theorem~\ref{theo:limit_set_convergence} asserts that the sequence $\{\bm{\theta}_k\}_{k\in\mathbb{N}}$ generated by the~\gls{gd} optimizer eventually approaches its limit point set $\Omega(\{\bm{\theta}_k\})$, on which $\Psi$ is constant. With Lemma~\ref{lemma:bounded_gradient}, this also implies that every point in the limit point set is critical. However, Theorem~\ref{theo:limit_set_convergence} does not guarantee convergence of the whole sequence $\{\bm{\theta}_k\}_{k\in\mathbb{N}}$ to a single point, since, in nonconvex problems, a bounded descent sequence may admit multiple accumulation points. To strengthen the result, we introduce a uniformized version of the~\gls{kl} property on $\Omega(\{\bm{\theta}_k\})$, which applies to compact nonempty sets on which the objective function is constant~\cite{attouch2013convergence,bolte2014proximal}, as is the case for $\Omega(\{\bm{\theta}_k\})$ by Theorem~\ref{theo:limit_set_convergence}.
    \begin{assumption}[Uniformized Kurdyka--{\L}ojasiewicz (K{\L}) property]
    \label{ass:kl}
        Let $\Omega(\{\bm{\theta}_k\})$ be the limit point set of the sequence generated by the~\gls{gd} scheme. There exist constants $\varepsilon>0$, $\eta>0$, and a continuous concave function $\varphi:[0,\eta)\to\mathbb{R}_+$ such that: 
        \begin{itemize}
            \item $\varphi(0)=0$, $\varphi$ is continuously differentiable on $(0,\eta)$, and $\varphi'(s)>0$ for all $s\in(0,\eta)$, with $\varphi'$ denoting the derivative of $\varphi$;
            \item for every $\bm{\theta}$ satisfying $\mathrm{dist}(\bm{\theta},\Omega(\{\bm{\theta}_k\})) < \varepsilon$ and $\Psi_{\infty} < \Psi(\bm{\theta}) < \Psi_{\infty}+\eta$, one has
            \begin{equation}
            \label{eq:uniform_kl}
                \varphi'(\Psi(\bm{\theta})-\Psi_{\infty})\,\|\nabla_{\bm{\theta}}\Psi(\bm{\theta})\| \ge 1.
            \end{equation}
        \end{itemize}
    \end{assumption}
    
    This property quantifies how the value of $\Psi$ controls the norm of its gradient in a neighborhood of the limit set $\Omega(\{\bm{\theta}_k\})$. It is \emph{uniformized} because the same constants $\varepsilon>0$, $\eta>0$, and the same concave function $\varphi$ apply uniformly to all points near $\Omega(\{\bm{\theta}_k\})$ satisfying $\Psi_{\infty}<\Psi(\bm{\theta})<\Psi_{\infty}+\eta$. The compactness of $\Omega(\{\bm{\theta}_{k}\})$ is essential, as shown in~\cite{attouch2013convergence,bolte2014proximal}, since it allows the pointwise~\gls{kl} property to be extended to the whole compact set, by covering it with finitely many neighborhoods. Assumption~\ref{ass:kl} holds whenever $\Psi$ satisfies the~\gls{kl} property at every point of $\Omega(\{\bm{\theta}_k\})$, since this set is compact and $\Psi$ is constant on it by Theorem~\ref{theo:limit_set_convergence}~\cite[Lemma~6]{bolte2014proximal}. The~\gls{kl} property, in turn, holds for real analytic functions~\cite{attouch2013convergence,bolte2014proximal}. In particular, if $\mathbf{f}$, $\mathbf{h}$, the costs, and $\tilde{\bm{\lambda}}$ are real analytic, then so is $\Psi$ on $\mathcal{U}_\theta$, by the analytic implicit function theorem; this is the case for the models and the parametrization used in Section~\ref{sec:numerical_experiments}.

    \begin{theorem}[Convergence of the~\gls{gd} scheme]
    \label{theo:total_convergence_gd}
        Let the assumptions of Lemma~\ref{lemma:sufficient_decrease} hold together with  Assumption~\ref{ass:kl}. Then, the sequence $\{\bm{\theta}_k\}_{k\in\mathbb{N}}$, generated by the~\gls{gd} scheme, satisfies the following statements.
        \begin{itemize}
            \item $\{\bm{\theta}_k\}_{k\in\mathbb{N}}$ is such that $\sum_{k=1}^{\infty} \|\bm{\theta}_{k} - \bm{\theta}_{k-1}\| < \infty$.
            \item $\{\bm{\theta}_k\}_{k\in\mathbb{N}}$ converges to a single limit point $\bm{\theta}_{*} \in \mathcal{L}_{0}$, which is a critical point of $\Psi$.
            \item If, in addition, $\Psi$ is twice continuously differentiable in a neighborhood of $\bm{\theta}_{*}$ and $\nabla_{\bm{\theta}}^{2}\Psi(\bm{\theta}_{*})\succ 0$, then $\bm{\theta}_{*}$ is a strict local minimizer of $\Psi$.
        \end{itemize}
    \end{theorem}
    \begin{proof}
        By Theorem~\ref{theo:limit_set_convergence}, the limit point set $\Omega(\{\bm{\theta}_k\})$ is nonempty and compact, $\Psi$ is constant on it with value $\Psi_{\infty}$, and $\operatorname{dist}(\bm{\theta}_k,\Omega(\{\bm{\theta}_k\})) \to 0$. Moreover, by Assumption~\ref{ass:kl}, there exist constants $\varepsilon>0$, $\eta>0$, and a function $\varphi$ satisfying~\eqref{eq:uniform_kl}. Since $\mathrm{dist}(\bm{\theta}_k,\Omega(\{\bm{\theta}_k\})) \to 0$ and $\Psi(\bm{\theta}_k)\to\Psi_\infty$, there exists $k_0\in\mathbb{N}$ such that, for every $k\ge k_0$, $\mathrm{dist}(\bm{\theta}_k,\Omega(\{\bm{\theta}_k\})) < \varepsilon$ and $\Psi_{\infty} \leq \Psi(\bm{\theta}_k) < \Psi_{\infty}+\eta$. In the case in which finite-time convergence in objective value occurs, i.e., $\Psi(\bm{\theta}_{k_f})=\Psi_{\infty}$ for some $k_f\in\mathbb{N}$, then the sufficient decrease property in Lemma~\ref{lemma:sufficient_decrease} implies $\bm{\theta}_{k+1}=\bm{\theta}_{k}$ for all $k\geq k_f$, and the result follows immediately. Otherwise, we can still assume that, for some $k_0$, $\Psi_{\infty} < \Psi(\bm{\theta}_k) < \Psi_{\infty}+\eta$ for all $k\geq k_0$. Hence, applying~\eqref{eq:uniform_kl} at $\bm{\theta}=\bm{\theta}_k$, we obtain 
        \begin{equation}
        \label{eq:uniform_kl_k}
            \varphi'(\Psi(\bm{\theta}_k)-\Psi_\infty)\,\|\nabla_{\bm{\theta}}\Psi(\bm{\theta}_k)\| \ge 1
            \qquad \forall k\ge k_0.
        \end{equation}
        From the~\gls{gd} update rule in Lemma~\ref{lemma:sufficient_decrease}
        \begin{equation}
        \label{eq:grad_increment}
            \|\nabla_{\bm{\theta}}\Psi(\bm{\theta}_{k})\| = \frac{\|\bm{\theta}_{k+1} - \bm{\theta}_{k}\|}{\alpha_{k+1}}.
        \end{equation}
        Substituting~\eqref{eq:grad_increment} in~\eqref{eq:uniform_kl_k} and using $\alpha_{k+1} \geq \underline{\alpha} > 0$ from the step-size condition in Lemma~\ref{lemma:sufficient_decrease},  $\forall k \geq k_{0}$ we obtain
        \begin{equation}
        \label{eq:varphi_and_theta}
            \varphi'(\Psi(\bm{\theta}_{k})-\Psi_{\infty}) \geq \frac{\alpha_{k+1}}{\|\bm{\theta}_{k+1} - \bm{\theta}_{k}\|} \geq \frac{\underline{\alpha}}{\|\bm{\theta}_{k+1} - \bm{\theta}_{k}\|}.
        \end{equation}
        Using the concavity of $\varphi$, we have
        \begin{equation}
        \label{eq:concavity}
            \varphi(a_{k}) - \varphi(a_{k+1}) \geq \varphi'(a_{k}) (a_{k}-a_{k+1}),
        \end{equation}
        where $a_{k} := \Psi(\bm{\theta}_{k}) - \Psi_{\infty}$ and $a_k-a_{k+1} = \Psi(\bm{\theta}_{k}) - \Psi(\bm{\theta}_{k+1})$. Since by the sufficient decrease result in Lemma~\ref{lemma:sufficient_decrease}
        \begin{equation}
        \label{eq:sufficient_decrease_proof}
            \Psi(\bm{\theta}_{k}) - \Psi(\bm{\theta}_{k+1}) \geq M \|\bm{\theta}_{k+1} - \bm{\theta}_{k}\|^{2}_{2}, 
        \end{equation}
        replacing~\eqref{eq:varphi_and_theta} and~\eqref{eq:sufficient_decrease_proof} in~\eqref{eq:concavity}, with simple manipulations we obtain
        \begin{equation}
            \|\bm{\theta}_{k+1} - \bm{\theta}_{k}\| \leq \frac{1}{M\underline{\alpha}} \left(\varphi(a_k) - \varphi(a_{k+1})\right) \qquad \forall k \geq k_{0}.
        \end{equation}
        Summing the above inequality for $k\ge k_{0}$ yields
        \begin{equation}
        \label{eq:k0_infty}
            \sum_{k=k_{0}}^{\infty} \|\bm{\theta}_{k+1} - \bm{\theta}_{k}\| \leq \frac{1}{M\underline{\alpha}} \varphi(\Psi(\bm{\theta}_{k_{0}}) - \Psi_{\infty})<+\infty
        \end{equation}
        since $\varphi(0) = 0$ and $\sum_{k=k_{0}}^{\infty}(\varphi(a_k) - \varphi(a_{k+1})) = \varphi(\Psi(\bm{\theta}_{k_{0}}) - \Psi_{\infty}) - \varphi(\Psi_{\infty} - \Psi_{\infty})$. Since the first $k_0$ terms are finite, $\sum_{k=0}^{\infty}\|\bm{\theta}_{k+1}-\bm{\theta}_{k}\|<+\infty$. Let $m,n\in\mathbb{N}$ be two arbitrary indices with $m>n$. Then, $\|\bm{\theta}_{m} - \bm{\theta}_{n}\| \leq \sum_{k=n}^{m-1} \|\bm{\theta}_{k+1} - \bm{\theta}_{k} \|$. Since the right-hand side tends to zero when $m, n \to \infty$, the sequence $\{\bm{\theta}_{k}\}_{k \in \mathbb{N}}$ is Cauchy and it converges to a single limit point $\bm{\theta}_{*}$. Finally, since all iterates belong to the closed set $\mathcal{L}_{0}$, it follows that $\bm{\theta}_{*}\in\mathcal{L}_{0}$. Moreover, $\bm{\theta}_{*}$ is an accumulation point of the sequence. Hence, by Lemma~\ref{lemma:bounded_gradient}, $\nabla_{\bm{\theta}}\Psi(\bm{\theta}_{*})=\mathbf{0}$. If $\Psi$ is twice continuously differentiable in a neighborhood of $\bm{\theta}_{*}$ and $\nabla_{\bm{\theta}}^{2}\Psi(\bm{\theta}_{*})\succ0$, then the standard second-order sufficient condition implies that $\bm{\theta}_{*}$ is a strict local minimizer of $\Psi$.
        \end{proof}
    Theorem~\ref{theo:total_convergence_gd} completes the local convergence analysis of~\gls{gd} applied to the reduced
    problem~\eqref{eq:reduced_min}. Under the stated assumptions, the iterates converge to a critical point of $\Psi$, which is a strict local minimizer under the second-order sufficient
    condition.
\end{document}

%% file: preamble/preamble.tex
\usepackage{amsmath,amssymb}
\usepackage{graphicx}
\usepackage{bm}
\usepackage{glossaries}
\usepackage{booktabs}
\usepackage{algorithm}
\usepackage{algpseudocode}
\theoremstyle{plain}
\newtheorem{theorem}{Theorem}
\newtheorem{lemma}[theorem]{Lemma}
\newtheorem{proposition}[theorem]{Proposition}
\newtheorem{corollary}[theorem]{Corollary}
\newtheorem{assumption}{Assumption}
\theoremstyle{definition}
\newtheorem{definition}{Definition}
\theoremstyle{remark}
\newtheorem{remark}{Remark}
\usepackage{etoolbox}
\apptocmd{\appendix}{%
    \setcounter{theorem}{0}%
    \setcounter{assumption}{0}%
    \setcounter{definition}{0}%
    \setcounter{remark}{0}%
    \setcounter{equation}{0}%
}{}{}

%% file: acronyms/acronyms.tex
\newacronym{kf}{KF}{Kalman Filter}
\newacronym{mhe}{MHE}{Moving Horizon Estimation}
\newacronym{mh}{MH}{Moving Horizon}
\newacronym{lti}{LTI}{Linear Time Invariant}
\newacronym{rl}{RL}{Reinforcement Learning}
\newacronym{dp}{DP}{Dynamic Programming}
\newacronym{rtss}{RTSS}{Rauch–Tung–Striebel Smoother}
\newacronym{map}{MAP}{Maximum A Posteriori}
\newacronym{fie}{FIE}{Full Information Estimation}
\newacronym{fnn}{FNN}{Feedforward Neural Network}
\newacronym{rnn}{RNN}{Recurrent Neural Network}
\newacronym{ekf}{EKF}{Extended Kalman Filter}
\newacronym{ukf}{UKF}{Unscented Kalman Filter}
\newacronym{ckf}{CKF}{Cubature Kalman Filter}
\newacronym{pf}{PF}{Particle Filter}
\newacronym{bf}{BF}{Bayes Filter}
\newacronym{pmp}{PMP}{Pontryagin's Maximum Principle}
\newacronym{nlp}{NLP}{Nonlinear Programming}
\newacronym{ertss}{ERTSS}{Extended Rauch–Tung–Striebel Smoother}
\newacronym{urtss}{URTSS}{Unscented Rauch–Tung–Striebel Smoother}
\newacronym{crtss}{CRTSS}{Cubature Rauch–Tung–Striebel Smoother}
\newacronym{bsps}{BSPS}{Backward-Simulation Particle Smoother}
\newacronym{mae}{MAE}{Mean Absolute Error}
\newacronym{nees}{NEES}{Normalized Estimation Error Squared}
\newacronym{ssosc}{SSOSC}{Strong Second-Order Sufficient Conditions}
\newacronym{palm}{PALM}{Proximal Alternating Linearized Minimization}
\newacronym{agd}{AGD}{Alternating Gradient Descent}
\newacronym{gd}{GD}{Gradient Descent}
\newacronym{kl}{KL}{Kurdyka--\L{}ojasiewicz}
\newacronym{wrt}{w.r.t.}{with respect to}
\newacronym{gelu}{GELU}{Gaussian Error Linear Unit}
\newacronym{td}{TD}{Temporal-Difference}